\documentclass[a4paper,11pt]{article}
\usepackage{jcappub} 
\usepackage{aas_macros}
\usepackage{lineno}
\newcommand{\tc}{\textcolor{black}}

\title{\boldmath Abundance and phase-space distribution of subhalos in cosmological N-body simulations: testing numerical convergence and correction methods}

\author{Kun Xu}
\affiliation{Center for Particle Cosmology, Department of Physics and Astronomy, University of Pennsylvania, Philadelphia, PA 19104, USA}

\emailAdd{kunxu@sas.upenn.edu}

\abstract{Subhalos play a crucial role in accurately modeling galaxy formation and galaxy-based cosmological probes within the highly nonlinear, virialized regime. However, numerical convergence of subhalo evolution is difficult to achieve, especially in the inner regions of host halos where tidal forces are strongest. I investigate the numerical convergence and correction methods for the abundance, spatial, and velocity distributions of subhalos using two $6144^3$-particle cosmological N-body simulations with different mass resolutions—\textsc{Jiutian-300} ($1.0 \times 10^{7}\,h^{-1}M_{\odot}$) and \textsc{Jiutian-1G} ($3.7 \times 10^{8}\,h^{-1}M_{\odot}$)—with subhalos identified by \textsc{hbt+}. My study shows that the Surviving subhalo Peak Mass Function (SPMF) converges only for subhalos with $m_{\rm peak}$ above 5000 particles but can be accurately recovered by including orphan subhalos that survive according to the merger timescale model of Jiang et al., which outperforms other models. Including orphan subhalos also enables recovery of the real-space spatial and velocity distributions to 5\%–10\% accuracy down to scales of 0.1–0.2$\,h^{-1}$Mpc. The remaining differences are likely due to cosmic variance and finite-box effects in the smaller \textsc{Jiutian-300} simulation. Convergence below 0.1$\,h^{-1}$Mpc remains challenging and requires more sophisticated modeling of orphan subhalos. I further highlight that redshift-space multipoles are more difficult to recover even at larger scales because unreliable small-scale pairs at $r_{\rm p} < 0.1\,h^{-1}$Mpc in real space affect scales of tens of Mpc in redshift space due to elongated Fingers-of-God effects. Therefore, for redshift-space statistics, I recommend using modified or alternative measures that reduce sensitivity to small projected separations in subhalo-based studies.}


\begin{document}
\maketitle
\flushbottom

\section{Introduction}
\label{sec:intro}
In the standard $\Lambda$ cold dark matter ($\Lambda$CDM) cosmological model, as well as in many of its modifications, structure forms hierarchically \cite{2012AnP...524..507F,1993MNRAS.262..627L}. Dark matter halos merge to form larger halos, with the smaller ones becoming substructures (subhalos). Galaxies form as gas cools and condenses at the centers of dark matter halos, while galaxies in smaller halos become satellite galaxies after mergers \cite{1978MNRAS.183..341W,1991ApJ...379...52W}. Therefore, substructures and their associated galaxies constitute an important component of the cosmic web. Understanding the formation and evolution of substructures is essential for a complete picture of cosmic structure formation and for accurately modeling galaxy formation and various cosmological probes. \tc{For example, a reliable subhalo model is essential for modeling galaxy clustering and abundances using methods such as Subhalo Abundance Matching \cite{2018ARA&A..56..435W}, as well as for studying satellite galaxy evolution in semi-analytical galaxy formation models \cite{2000MNRAS.319..168C}. In cosmology, a robust subhalo model is also crucial for interpolating the strong-lensing probability \cite{2016ApJ...819..114K,2020Sci...369.1347M} and dark matter annihilation rate \cite{2012MNRAS.427.1651H}, and accurate subhalo spatial and velocity distributions are key to modeling the contribution of satellite kinematics to redshift-space distortions \cite{2012ApJ...758...50L}.}

After infall, subhalos are primarily affected by dynamical friction \cite{1943ApJ....97..255C}, tidal stripping, and tidal heating \cite{1997ApJ...474..223G}, until they either merge with the central subhalo or are completely disrupted. Some studies have argued that the central cusp of a Navarro–Frenk–White (NFW) profile \cite{1997ApJ...490..493N} subhalo is expected to survive tidal forces in most cases, even after intense and prolonged stripping \cite{2010MNRAS.402...21N}. The nonlinearity and chaotic evolution of subhalos within virialized host systems necessitate the use of semi-analytic \cite{2001ApJ...559..716T, 2022MNRAS.510.2900D} or numerical \cite{2001MNRAS.328..726S, 2004MNRAS.355..819G,2012MNRAS.427.2437H, 2013ApJ...762..109B} methods to model their dynamics. However, recent work \cite{2018MNRAS.475.4066V} has demonstrated that numerical limitations significantly influence subhalo evolution, artificially enhancing tidal disruption and impeding convergence in subhalo population statistics. Without properly accounting for subhalos affected by numerical disruption, simulations will underestimate their abundance by missing a significant fraction of the population \tc{that has been entirely artificially destroyed.} \cite{2004MNRAS.355..819G,2016MNRAS.457.1208H}. A commonly adopted approach to recover the lost population is to trace their orbits using the most bound particle prior to disruption \cite{2016MNRAS.457.1208H,2025MNRAS.540.1107S} or to model their subsequent trajectories semi-analytically \cite{2022MNRAS.510.2900D}. Subhalos and the galaxies they host in such cases are typically referred to as``type-2 subhalos” and ``orphans”.

It has recently been shown that subhalos that fall in before $z = 2$ and higher-order subhalos (i.e., sub-subhalos) experience extreme mass loss, are more susceptible to numerical disruption, and constitute the majority of orphan subhalos \cite{2025ApJ...981..108H}. \tc{Nevertheless, most of their mass loss is physical rather than purely numerical, so in reality the surviving systems primarily populate the low-mass end of the present-day subhalo population.} Therefore, for studies related to the {\it current} mass of subhalos—such as gravitational lensing \cite{2016ApJ...819..114K} and dark matter annihilation \cite{2012MNRAS.427.1651H}—\tc{ neglecting the contribution from orphan subhalos may have only a minor impact.} However, in galaxy-related studies, stellar mass and luminosity show the strongest correlation with the maximum mass ever attained by a subhalo ($m_{\rm peak}$) or its highest circular velocity ($V_{\rm peak}$) \cite{2018ARA&A..56..435W}. This relationship exists because satellite galaxies experience mostly passive evolution after falling into their host system, and their stars are significantly more resistant to tidal stripping compared to the dark matter halos that surround them \cite{2016ApJ...833..109S}. Consequently, \tc{neglecting subhalos that are entirely artificially disrupted} can impact subhalos hosting massive galaxies—typically observed in current galaxy surveys—and should be treated carefully.

Orphans are typically tracked by following the trajectories of their most bound particles prior to disruption, with their survival determined using merger timescales from dynamical friction models \cite{1993MNRAS.262..627L,2008MNRAS.383...93B,2008ApJ...675.1095J}. This can be regarded as a leading-order correction, since the evolution of resolved subhalos can differ from that of individual particles, resulting in distinct phase-space distributions. Although more sophisticated semi-analytical approaches exist \cite{2001ApJ...559..716T,2022MNRAS.510.2900D}, no widely accepted next-level treatment has yet emerged. The above method has already been adopted in empirical and semi-analytical galaxy formation models \cite{2000MNRAS.319..168C,2010MNRAS.404.1111G,2019MNRAS.488.3143B}. However, a comprehensive evaluation of these correction methods in cosmological simulations has not yet been performed, with only limited examinations in zoom-in simulations of individual Milky Way–like halos \cite{2016MNRAS.457.1208H,2025MNRAS.540.1107S}. In this study, I compare subhalo abundance and phase-space distributions in two high-resolution cosmological simulations with different mass resolutions. I find that the surviving subhalo peak mass function converges only for subhalos with $m_{\rm peak}$ exceeding 5000 particles, but can be accurately recovered by including orphan subhalos that survive according to the merger timescale model of Jiang et al. \cite{2008ApJ...675.1095J}, which outperforms alternative models. Including orphans also enables recovery of the real-space spatial and velocity distributions to 5–10\% accuracy down to scales of 0.1–0.2$\,h^{-1}$Mpc, though convergence below 0.1$\,h^{-1}$Mpc remains challenging and likely requires more sophisticated orphan modeling. These results also contribute to the study and understanding of these fundamental properties of subhalos.

The paper is organized as follows. Section~\ref{sec:data} describes the simulations used in this work. The subhalo abundance, spatial distribution, and velocity distribution are analyzed in sections~\ref{sec:pmf}, \ref{sec:space_dis}, and \ref{sec:velocity}, respectively. A summary is provided in section~\ref{sec:con}.

\section{Simulation data}\label{sec:data}
In this work, I employ the primary runs of the \textsc{Jiutian} hybrid simulation suite \cite{2025arXiv250321368H}, developed for the extragalactic surveys of the China Space Station Telescope (CSST) \cite{2019ApJ...883..203G}. The suite comprises three primary high-resolution $\Lambda$CDM simulations based on Planck 2018 cosmology \cite{2020A&A...641A...6P}, 129 medium-resolution $\nu w_0 w_a$CDM simulations from the \textsc{Kun} emulator \cite{2025SCPMA..6889512C}, nine high-resolution zoom-in hydrodynamical simulations centered on the Milky Way with initial conditions reconstructed by the \textsc{ELUCID} project \cite{2014ApJ...794...94W}, and 21 additional runs exploring extensions including warm DM \cite{2023MNRAS.526.3156H}, $f(R)$ gravity \cite{2012JCAP...01..051L}, interacting dark energy \cite{2018PhRvD..98j3530Z}, and $\nu\Lambda$CDM cosmologies. 

The three primary $\Lambda$CDM runs are dark matter-only simulations performed with $6144^3$ particles, differing in box size $L$ and particle mass resolution $m_{\rm p}$. \textsc{Jiutian-300} has $L = 300\,h^{-1}\mathrm{Mpc}$ and $m_{\rm p} = 1.005 \times 10^7\,h^{-1}M_\odot$; \textsc{Jiutian-1G} has $L = 1\,h^{-1}\mathrm{Gpc}$ and $m_{\rm p} = 3.723 \times 10^8\,h^{-1}M_\odot$; and \textsc{Jiutian-2G} has $L = 2\,h^{-1}\mathrm{Gpc}$ and $m_{\rm p} = 2.978 \times 10^9\,h^{-1}M_\odot$. The simulations were carried out using either the \textsc{Gadget-3} \cite{2012MNRAS.426.2046A} or \textsc{Gadget-4} \cite{2021MNRAS.506.2871S} codes. DM halos are identified with the friends-of-friends (FOF) algorithm \cite{1985ApJ...292..371D}, adopting a linking length of 0.2 times the mean inter-particle separation. 

Subhalos and merger trees are constructed using \textsc{hbt+} \cite{2012MNRAS.427.2437H,2018MNRAS.474..604H}, a time-domain subhalo finder and merger tree builder. The algorithm traces the evolution of each halo across simulation snapshots, from early times to the present. When a halo merges into a more massive one, it becomes a subhalo, tracked via its self-bound particles. Even after a subhalo becomes unresolved, \textsc{hbt+} continues to track its most bound particles—referred to as an ``orphan"—thereby preserving its identity. This approach naturally avoids many of the challenges faced by traditional configuration-space subhalo finders and yields physically consistent subhalo catalogs along with their full merger histories. 

In this study, I make use of the \textsc{Jiutian-300} and \textsc{Jiutian-1G} simulations.

\begin{figure}
    \centering
    \includegraphics[width=\textwidth]{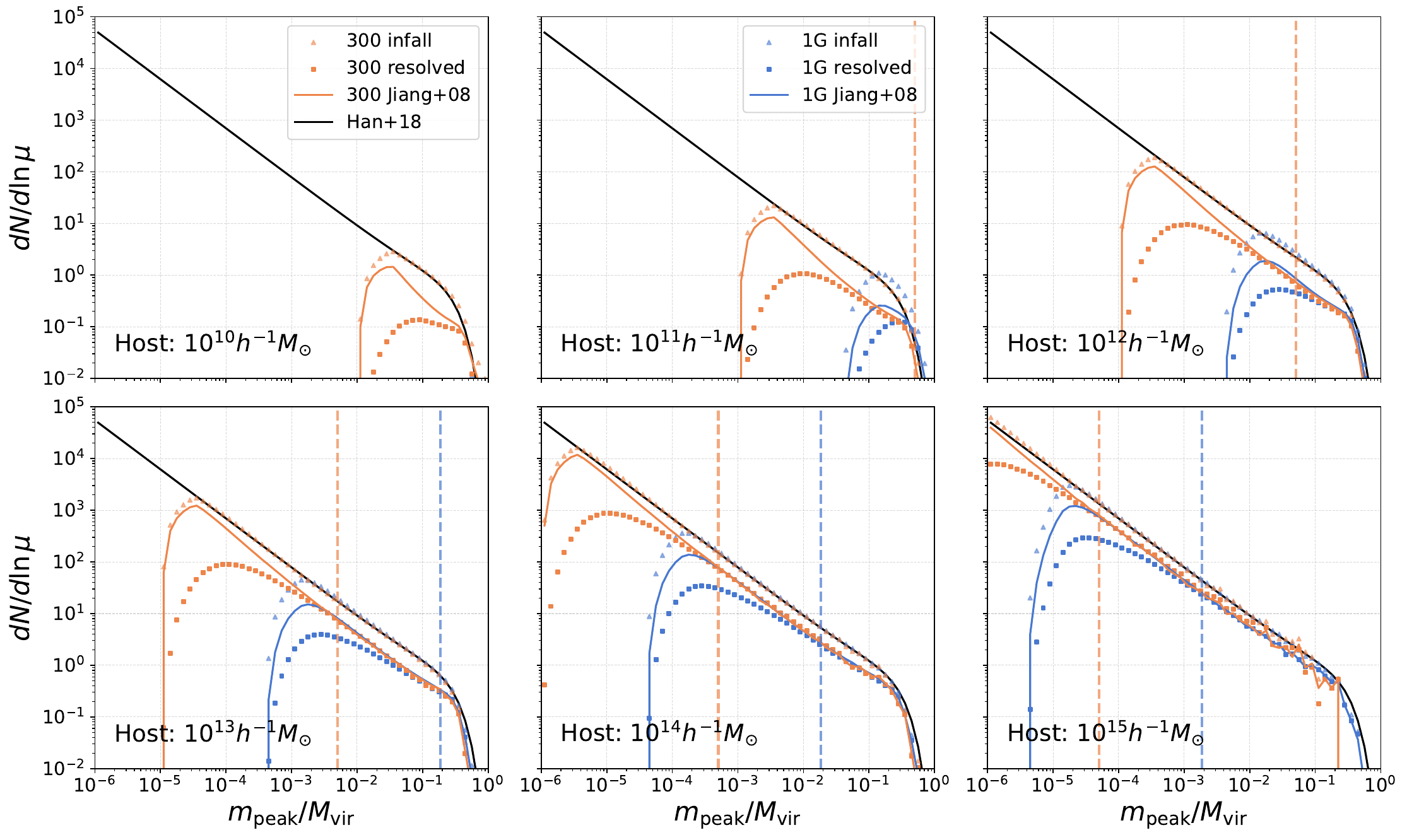}
    \caption{ Subhalo peak mass functions of surviving subhalos (squares) and all infall subhalos (triangles) in \textsc{Jiutian-300} (orange) and \textsc{Jiutian-1G} (blue) within $R_{\rm vir}$ at $z=0$, shown for host halos in different mass bins. Black solid lines indicate the universal IPMFs from Han et al. \cite{2018MNRAS.474..604H}. The orange and blue solid lines show the SPMFs predicted by applying the merger timescale model of Jiang et al. \cite{2008ApJ...675.1095J} to all infall subhalos in \textsc{Jiutian-300} and \textsc{Jiutian-1G}, regardless of their present-day status. Dashed vertical lines indicate the subhalo mass corresponding to 5000 particles.}
    \label{fig:SPMF_HM}
\end{figure}

\section{Abundance of surviving subhalos}\label{sec:pmf}
In this section, I investigate the numerical convergence of the Surviving subhalo Peak Mass Function (SPMF) using the \textsc{Jiutian-1G} and \textsc{Jiutian-300} simulations. I test commonly used correction methods for subhalo abundance loss due to numerical disruption and propose an improved version. Additionally, I examine the dependence of the SPMF on host halo mass and its redshift evolution, and provide an accurate fitting formula.

\subsection{Numerical convergence and correction methods}
I test the convergence of the SPMF by comparing results from the \textsc{Jiutian-300} and \textsc{Jiutian-1G} simulations. Subhalos are considered resolved if they contain more than 20 bound particles at $z = 0$, a commonly adopted threshold in many simulations. I count the number of subhalos within $R_{\rm vir}$ as a function of the peak mass ratio $\mu = m_{\rm peak} / M_{\rm vir}$ within each $M_{\rm vir}$ bin, and normalize it by the number of host halos in the same bin. Here, $M_{\rm vir}$ and $R_{\rm vir}$ denote the present-day virial mass and radius of the host halo, and $m_{\rm peak}$ is the maximum bound mass attained by the subhalo over its history. $M_{\rm vir}$ and $R_{\rm vir}$ is defined in a redshift-dependent manner based on the spherical collapse model, with the overdensity criterion from Bryan \& Norman \cite{1998ApJ...495...80B}. 

I present the SPMFs for different host halo mass bins in Figure~\ref{fig:SPMF_HM}, using logarithmic intervals of width $10^{0.5}$ centered on the values indicated in each panel. The results for \textsc{Jiutian-300} and \textsc{Jiutian-1G} are presented in orange and blue squares respectively. By comparing the results from the two simulations, I find that the SPMF converges only down to peak subhalo masses corresponding to approximately 5000 particles ($10^{10.7}\,h^{-1}M_{\odot}$ for \textsc{Jiutian-300} and $10^{12.3}\,h^{-1}M_{\odot}$ for \textsc{Jiutian-1G}), as indicated by the vertical dashed line. This threshold is consistent with the value reported in ref.~\cite{2024ApJ...970..178M}, which found convergence at around 4000 particles. The SPMFs from \textsc{Jiutian-1G} increasingly deviate from those of \textsc{Jiutian-300} toward the low-mass end, indicating that subhalos with fewer particles are more susceptible to numerical disruption \cite{2018MNRAS.475.4066V,2025ApJ...981..108H}.

\begin{figure}
    \centering
    \includegraphics[width=\textwidth]{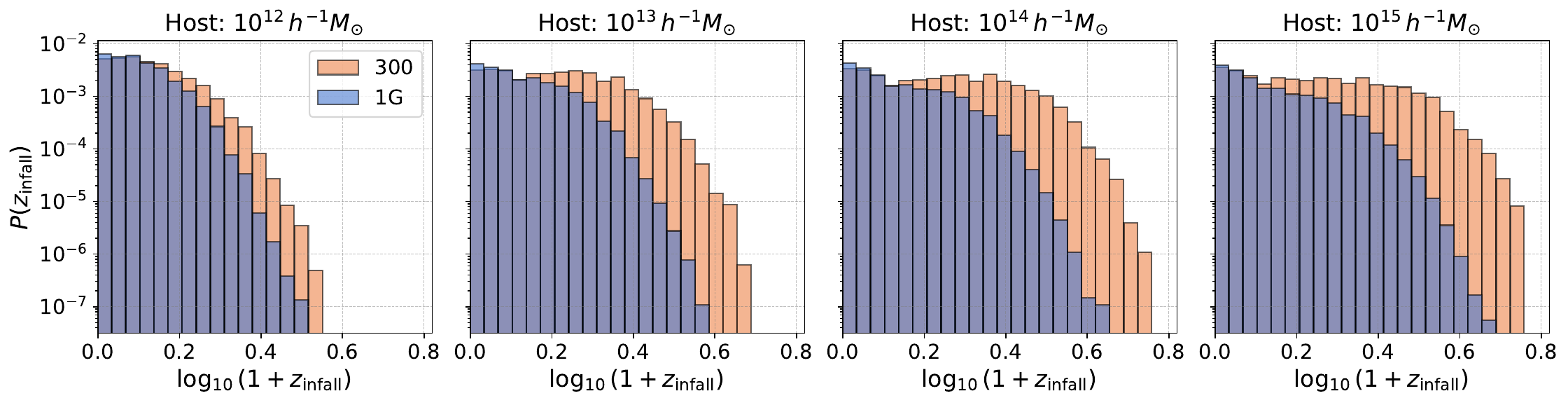}
    \caption{Infall redshift distributions of resolved subhalos with peak masses in the range $10^{10.7} < m_{\rm peak} < 10^{12.3}\,h^{-1}M_{\odot}$. In this range, subhalos are fully resolved in \textsc{Jiutian-300} but are affected by numerical effects in \textsc{Jiutian-1G}. Results are shown for different host halo mass bins. The \textsc{Jiutian-1G} distributions are normalized using the subhalo counts from \textsc{Jiutian-300}, scaled by the simulation volumes. As a result, the histogram area for \textsc{Jiutian-300} is 1, while that for \textsc{Jiutian-1G} is less than 1.}
    \label{fig:zdis}
\end{figure}

He et al. \cite{2025ApJ...981..108H} found that subhalos which fall in before $z = 2$ and are higher-order subhalos (i.e., sub-subhalos) experience extreme mass loss and are more susceptible to disruption, either physically or numerically. These subhalos undergo preprocessing and are often accreted through major mergers at high redshift, constituting the majority of orphan subhalos. To confirm this result, I show the infall redshift ($z_{\rm infall}$) distributions of resolved subhalos with peak masses in the range $10^{10.7} < m_{\rm peak} < 10^{12.3}\,h^{-1}M_{\odot}$ in figure~\ref{fig:zdis}. Here, $z_{\rm infall}$ is defined as the redshift when the subhalo first crosses the virial radius $R_{\rm vir}(z_{\rm infall})$ of its host halo. Within this mass range, subhalos are fully resolved in \textsc{Jiutian-300} but are affected by numerical effects in \textsc{Jiutian-1G}. The distribution from \textsc{Jiutian-1G} is normalized by the subhalo counts in \textsc{Jiutian-300}, scaled by the simulation volumes, to better highlight the differences. Figure~\ref{fig:zdis} shows that most of the numerically disrupted subhalos in \textsc{Jiutian-1G} have $z_{\rm infall} > 1$ across all host halo mass bins, with the disrupted fraction increasing with $z_{\rm infall}$.

To recover subhalo abundance lost to numerical disruption, \textsc{hbt+} tracks the most bound particles of unresolved orphan subhalos. However, including all orphans can lead to an overcorrection of the subhalo abundance, as the tracked particles do not experience dynamical friction \cite{1943ApJ....97..255C}, but instead follow thermal motions similar to those of the surrounding diffuse particles, given their identical mass. As a result, these orphans may artificially survive indefinitely and never merge with the central subhalo. In reality—or in higher-resolution simulations where such subhalos are resolved—they remain significantly more massive than background particles and experience dynamical friction due to frequent two-body interactions. Therefore, a commonly adopted approach is to calculate the merger timescale based on the infall conditions of subhalos—such as their mass ratio and orbital parameters—at a stage when they are not yet significantly affected by numerical disruption \cite{1993MNRAS.262..627L,2008MNRAS.383...93B,2008ApJ...675.1095J,2017MNRAS.472.1392S,2025ApJ...986..201X}. An orphan subhalo is then considered merged once its survival time exceeds the predicted merger timescale.

In figure~\ref{fig:SPMF_HM}, I show the infall subhalo peak mass functions (IPMFs) from \textsc{Jiutian-300} and \textsc{Jiutian-1G} using triangle markers, and compare them with the universal model from Han et al.~\cite{2018MNRAS.474..604H}. The IPMFs are obtained by counting all infall subhalos, regardless of whether they are surviving, merged, or disrupted by $z=0$. The IPMFs from \textsc{Jiutian-300} and \textsc{Jiutian-1G} show good convergence and agree well with the universal model of Han et al., confirming the universality of the subhalo mass function. Recently, it has been shown that this universality originates from a more fundamental quantity—the universal specific merger rate of dark matter halos \cite{2022ApJ...929..120D,2025arXiv250220181J}. The difference between the IPMF and SPMF indicates the extent to which subhalos have merged or been disrupted after infall, which must be carefully accounted for using merger timescale models.

Although this method can, in principle, recover the correct subhalo abundance, the phase-space distribution of orphan subhalos remains inaccurate \cite{2025MNRAS.540.1107S}. Orphans follow the same distribution as the surrounding dark matter particles, whereas, if resolved, they would possess lower energy and angular momentum due to dynamical friction, resulting in a distinct phase-space structure. It has been shown that resolved subhalos exhibit a more cuspy inner spatial distribution than the dark matter density profile, with more massive subhalos displaying steeper profiles—a consequence of dynamical friction \cite{2016MNRAS.457.1208H}. This will also be tested in section \ref{sec:space_dis}. More accurate recovery of the spatial or phase-space distribution requires further empirical corrections or semi-analytical modeling of the evolution of orphan subhalos \cite{2001ApJ...559..716T,2022MNRAS.510.2900D}.

\begin{figure}
    \centering
    \includegraphics[width=\textwidth]{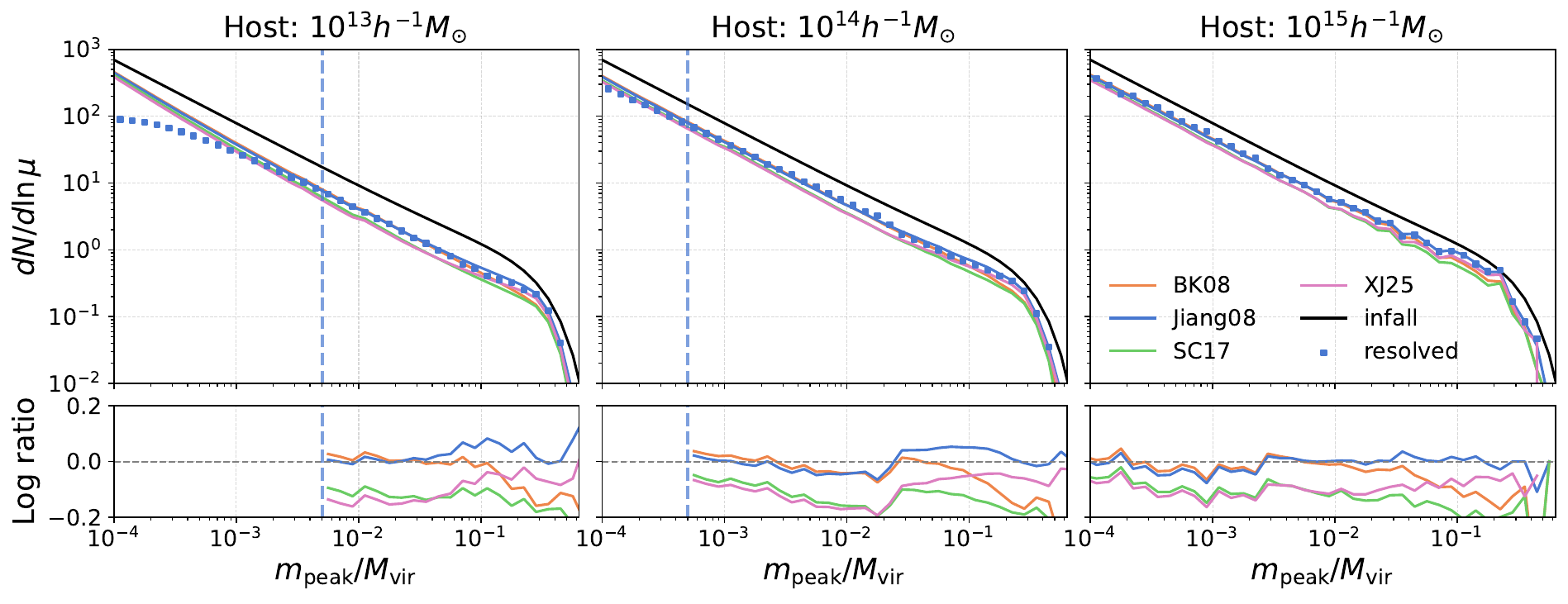}
    \caption{Comparison of SPMFs within $R_{\rm vir}$ from the combination of \textsc{Jiutian-300} and \textsc{Jiutian-1G} at $z=0$ with predictions obtained by checking whether each infall subhalo should have merged according to the merger timescale $T_{\rm merger}$ from four different models. Results are shown for three host halo mass bins. The blue dashed vertical lines indicate the subhalo mass corresponding to 5000 particles in \textsc{Jiutian-300}. The universal IPMF from Han et al. \cite{2018MNRAS.474..604H} is shown as black solid lines for comparison.}
    \label{fig:compare}
\end{figure}

\subsection{Comparison of merger timescale models}
Various models exist for estimating merger timescales, but they typically take the form of 
\begin{equation}  
\frac{T_{\rm{merger}}}{T_{\rm{dyn}}} = A\frac{f(\epsilon)}{\ln\Lambda} \left[\frac{r_c(E)}{r_{\rm{host}}}\right]^{\gamma_t} \left[\frac{M_{\rm{host}}}{M_{\rm{sat}}}\right]^{\beta_t}\,.  \label{eq:mergert}
\end{equation}
The parameter $\epsilon = j_{\rm{infall}} / j_c(E)$ characterizes the orbital circularity, where $r_c(E)$ and $j_c(E)$ are the radius and specific angular momentum of a circular orbit with the same energy $E$. The host halo mass $M_{\rm{host}}$ and radius $r_{\rm{host}}$ correspond to the virial mass $M_{\rm{vir}}$ and virial radius $r_{\rm{vir}}$ at $z_{\rm infall}$, respectively, and $M_{\rm{sat}}$ is the subhalo bound mass. The dynamical time is defined as $T_{\rm{dyn}} = \sqrt{r_{\rm{host}}^3 / (G M_{\rm{host}})}$, and $\ln\Lambda$ is the standard Coulomb logarithm, typically with $\Lambda = 1 + M_{\rm{host}} / M_{\rm{sat}}$.

Here, I consider four models for merger timescales. Using idealized N-body simulations, Boylan-Kolchin et al. \cite{2008MNRAS.383...93B} (BK08) found that $f(\epsilon) = e^{1.9\epsilon}$, with parameters $\beta_t = 1.3$ and $\gamma_t = 1.0$. In contrast, results from cosmological hydrodynamic simulations by Jiang et al. \cite{2008ApJ...675.1095J} (Jiang08) suggest a much weaker dependence of merger timescales on the infall orbital parameters, with $f(\epsilon) = 0.90\epsilon^{0.47} + 0.6$, $\beta_t = 1.0$, and $\gamma_t = 0.5$. Furthermore, Simha \& Cole \cite{2017MNRAS.472.1392S} (SC17), using cosmological N-body simulations, obtained results consistent with the earlier findings of Lacey \& Cole \cite{1993MNRAS.262..627L}, finding $f(\epsilon) = \epsilon^{0.85}$ with $\beta_t = 1.0$ and $\gamma_t = 1.8$. More recently, Xu \& Jing \cite{2025ApJ...986..201X} (XJ25), using the IllustrisTNG hydrodynamic simulations \cite{2018MNRAS.473.4077P}, found that merger timescales are not universal across different systems. They proposed a model of the form $f(\epsilon) = e^{\alpha_t\epsilon}$, where $\alpha_t$, $\beta_t$, and $\gamma_t$ vary with halo mass. Their results show that in lower-mass host halos, merger timescales have a stronger dependence on infall conditions, resembling BK08, while in higher-mass halos, the dependence weakens and approaches the behavior reported in Jiang08.

\begin{figure}
    \centering
    \includegraphics[width=\textwidth]{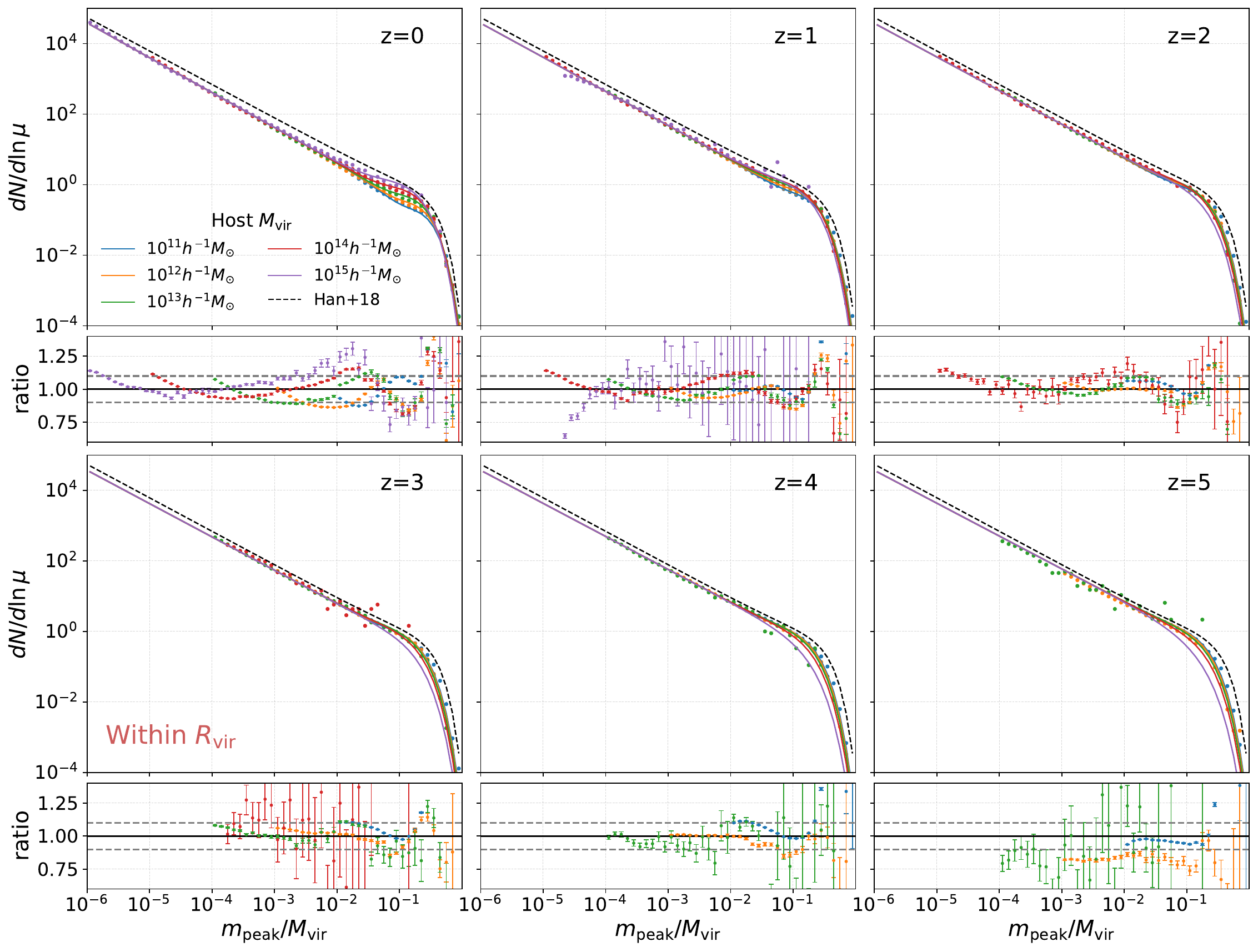}
    \caption{Host halo mass dependence and redshift evolution of the SPMF within $R_{\rm vir}$. Dots represent results from simulations, while solid lines show the fitting formula. The universal IPMF from Han et al. \cite{2018MNRAS.474..604H} is shown as black dashed lines for comparison. Dashed horizontal lines in the ratio panels indicate the $\pm10\%$ deviation range. Poisson noise is shown as error bars.}
    \label{fig:evo_rvir}
\end{figure}

\begin{figure}
    \centering
    \includegraphics[width=\textwidth]{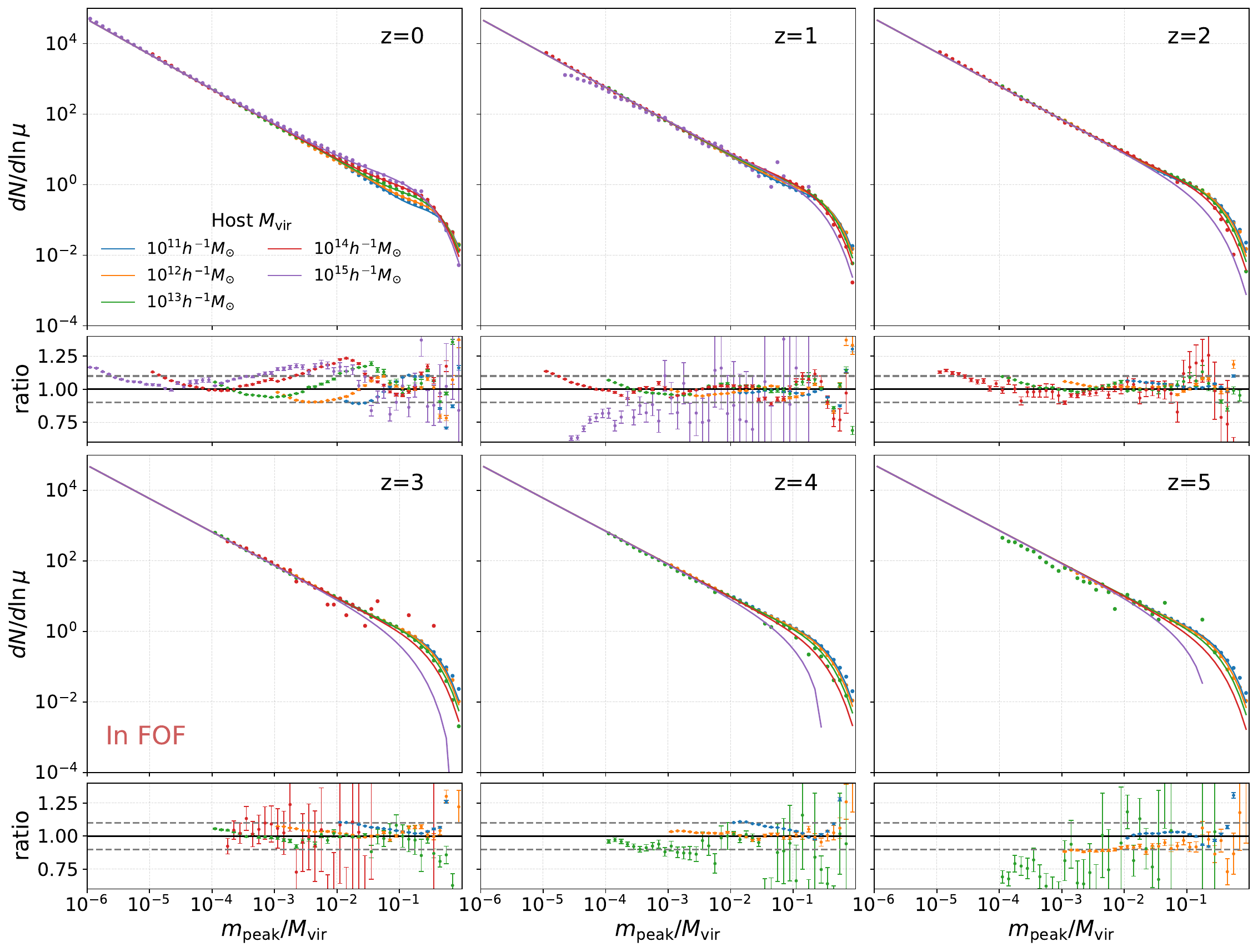}
    \caption{Same as figure~\ref{fig:evo_rvir}, but counting all subhalos within the FOF groups instead of only those inside $R_{\rm vir}$.}
    \label{fig:evo_fof}
\end{figure}

I directly test these models by comparing the SPMFs in mass ranges that are fully resolved in \textsc{Jiutian-300} or \textsc{Jiutian-1G} to the results obtained by checking whether each infall subhalo should have merged according to the merger timescale $T_{\rm merger}$ from these models. Here, I do not consider the current state of the subhalos—i.e., even subhalos that are still resolved may be identified as already merged. The results from an accurate model should closely reproduce the SPMFs from the simulations in these mass ranges. The comparison is shown in figure~\ref{fig:compare}. For the resolved SPMFs from simulations, I combine the results from \textsc{Jiutian-300} and \textsc{Jiutian-1G}, using the high-mass end from \textsc{Jiutian-1G}'s resolved range and \textsc{Jiutian-300} otherwise. This approach minimizes the impact of cosmic variance and finite box size effects. For the XJ25 model, I adopt the parameterization from the first row of their table 1. For higher-order subhalos, $T_{\rm merger}$ is calculated relative to their immediate host (i.e., the subhalo hosting the sub-subhalo), which has been shown to be a good leading-order approximation \cite{2010A&A...510A..60J}. 

Figure~\ref{fig:compare} shows that the Jiang08 model reproduces the SPMFs quite well across the entire mass ratio range, with only a slight overestimation at $10^{-1.5} < \mu < 10^{-0.5}$ for low-mass host halos. The BK08 model performs well at the low mass ratio end but significantly underestimates the abundance at high mass ratios. Both SC17 and XJ25 underestimate the entire range, though XJ25 provides a better match at the high mass ratio end. In principle, XJ25 should perform best, as it is based on state-of-the-art simulations and incorporates second-order effects in its model. I find that this discrepancy arises from differences in the subhalo finders. IllustrisTNG uses the configuration-space-based subhalo finder \textsc{subfind} \cite{2001MNRAS.328..726S}, which has been shown to systematically underestimate subhalo masses even at infall (see figure~B1 in ref.~\cite{2025arXiv250206932F}), whereas \textsc{hbt+} has been demonstrated to be more robust. Since the XJ25 model is calibrated using the underestimated subhalo masses from \textsc{subfind}, applying it to \textsc{hbt+} subhalo masses can lead to an underestimation of $T_{\rm merger}$, and consequently, the subhalo abundance. The subhalos used in SC17 are also identified with \textsc{subfind}, leading to similar biases. These models may still be useful for handling orphans in \textsc{subfind}-based simulations, which are subject to the same systematics. Although the Jiang08 model predates the formal introduction of \textsc{hbt+} \cite{2012MNRAS.427.2437H}, it adopts a similar time-domain approach to identify subhalos. The consistency between the Jiang08 results and the BK08 results at low mass ratios supports the use of time-domain subhalo finders like \textsc{hbt+}. In contrast, BK08 fails at high mass ratios due to the limitations of its idealized setup. A more accurate merger timescale model could be developed by applying the XJ25 method to catalogs produced by time-domain subhalo finders. For better comparison, I also include the results from the Jiang08 model in figure~\ref{fig:SPMF_HM}. 

\subsection{Host halo mass dependence and evolution}
By combining the resolved SPMFs from \textsc{Jiutian-300} and \textsc{Jiutian-1G}, and applying the Jiang08 model to correct for low mass ratios, I examine the host halo mass dependence and evolution of the SPMF. This offers a largely model-independent approach, as only the lowest mass ratio end is affected by empirical modeling.

Figure~\ref{fig:evo_rvir} shows the SPMFs within $R_{\rm vir}$ for different host halo mass bins at redshifts $z = 0, 1, 2, 3, 4,$ and $5$. While using $R_{\rm vir}$ is convenient for comparison with theoretical models—since it represents a natural boundary from spherical collapse theory—subhalos are often defined within the full FOF group in simulations, which typically extends beyond $R_{\rm vir}$. For practical empirical applications, I also present the SPMFs measured within FOF groups in figure~\ref{fig:evo_fof}. At each redshift—particularly at lower ones—the SPMFs are higher in more massive host halos at large $\mu$, while they tend to agree at small $\mu$. This is due to the difference of infall redshift distributions of the subhalos \cite{2011ApJ...741...13Y}. At fixed $\mu$, $m_{\rm peak}$ is smaller in low mass host halos, which formed earlier and also infall earlier, see figure 7 of ref. \cite{2011ApJ...741...13Y}. They have more time to merge with the central subhalo, and the merger timescale $T_{\rm merger}$ is shorter as the dynamical time $T_{\rm dyn} \propto (G\rho_c)^{-1/2}$ in eq.~\eqref{eq:mergert} is smaller at earlier times. These combined effects result in the host halo mass dependence at large $\mu$. For smaller $\mu$, the merger timescales $T_{\rm merger}$ of these subhalos are generally long enough that the differences become less significant.

It has been shown that subhalo mass functions, under various definitions, can all be well described by a double Schechter form \cite{2018MNRAS.474..604H}. I find that this also holds true for the SPMFs. Following Han et al. \cite{2018MNRAS.474..604H}, I model the SPMFs using a double Schechter function of the form
\begin{equation}
    \frac{{\rm d}N}{{\rm d}\ln\mu}=\left(a \mu^{-\alpha} + b \mu^{-\beta}\right) \exp\left(-c \mu^{\gamma}\right)\label{eq:SHMF_schechter}\,,
\end{equation}
where $\alpha$ describes the slope at the low-$\mu$ end. To capture the host halo mass dependence at large $\mu$, I find it sufficient to parameterize $b$ and $\beta$ as functions of $M_{\rm vir}$:
\begin{equation}
b = b_{m} + b_n \log_{10}\left(\frac{M_{\rm vir}}{10^{12}\,h^{-1}M_{\odot}}\right), \quad \beta = \beta_{m} + \beta_n \log_{10}\left(\frac{M_{\rm vir}}{10^{12}\,h^{-1}M_{\odot}}\right)\,, \label{eq:SHMF_mass}
\end{equation}
Moreover, I find that $c$ and $\gamma$ remain nearly constant across different redshifts. To capture the redshift evolution, I parameterize $a$, $\alpha$, $b_m$, $b_n$, $\beta_m$, and $\beta_n$ as power laws of $(1+z)$:
\begin{equation}
\label{eq:SHMF_z}
\begin{alignedat}{3}
a        &= d_{a}(1+z)^{\eta_a}            &\qquad\quad
\alpha   &= d_{\alpha}(1+z)^{\eta_\alpha}  &\qquad\quad
b_m      &= d_b^m(1+z)^{\eta_b^m} \,, \\
b_n      &= -d_b^n(1+z)^{\eta_b^n}         &\qquad
\beta_m  &= -d_\beta^m(1+z)^{\eta_\beta^m} &\qquad
\beta_n  &= d_\beta^n(1+z)^{\eta_\beta^n} \,,
\end{alignedat}
\end{equation}

\begin{table}[htbp]
\centering
\resizebox{\textwidth}{!}{ 
\begin{tabular}{ll|cccccccccccccc}
\hline
\hline
Type & Region 
& $d_{a}$ & $\eta_a$ & $d_{\alpha}$ & $\eta_\alpha$ & $d_b^m$ & $\eta_b^m$ 
& $d_b^n$ & $\eta_b^n$ & $d_\beta^m$ & $\eta_\beta^m$ & $d_\beta^n$ & $\eta_\beta^n$ & $c$ & $\gamma$ \\
\hline
IPMF 
& $R_{\rm vir}$
& 0.11& -- & 0.95 &  -- & 0.32 &  
& --  &  -- & 0.08 &  -- &  -- & -- & 8.9 & 1.9\\

SPMF 
& $R_{\rm vir}$
& 0.046& 0.41& 0.99 & -0.032 & 26 & -0.99 
& 5.1 & -0.73 & 1.7 & -0.75 & 0.34 &-0.66 & 14 & 1.2\\

SPMF 
& FOF 
& 0.054& 0.52& 0.99 & -0.036 & 3.9 & -0.51 
& 0.83 & -0.13 & 1.0 & -1.2 & 0.37 &-0.65 &6.1 & 1.0\\
\hline
\end{tabular}
}
\caption{Parameters of the subhalo PMF fitting function. IPMF results are taken from Han et al. \cite{2018MNRAS.474..604H}.}
\label{tab:spmf_params}
\end{table}

I fit the SPMFs across 86 snapshots within $0<z<5$, in 10 host halo mass bins spanning $10^{10.0}<M_{\rm vir}<10^{15.5}\,h^{-1}M_{\odot}$, and over the range $10^{-6}<\mu<1$, providing separate models for subhalos within $R_{\rm vir}$ and those identified in FOF. I list the best-fit parameters in table~\ref{tab:spmf_params}, and plot the best-fit models in figure~\ref{fig:evo_rvir} and figure~\ref{fig:evo_fof}. My models achieve an overall accuracy of 10\% across all redshifts, $M_{\rm vir}$, and $\mu$ bins. This model is useful for empirical applications that rely on the SPMF, and also serves as a good starting point for extended studies, such as modeling the subhalo mass function of current mass (subhalo evolved mass function), which can be obtained by combining it with a tidal stripping model \cite{2016MNRAS.457.1208H}. He et al. \cite{2025ApJ...981..108H} found that the evolved mass function is largely insensitive to numerical disruption and the treatment of orphans, as these subhalos experience significant mass loss and contribute mainly to the very low-$\mu$ end. Therefore, I do not investigate the evolved mass function here and refer interested readers to Han et al. \cite{2018MNRAS.474..604H} for a well-established model.  

\begin{figure}[htbp]
    \centering
    \includegraphics[width=\textwidth]{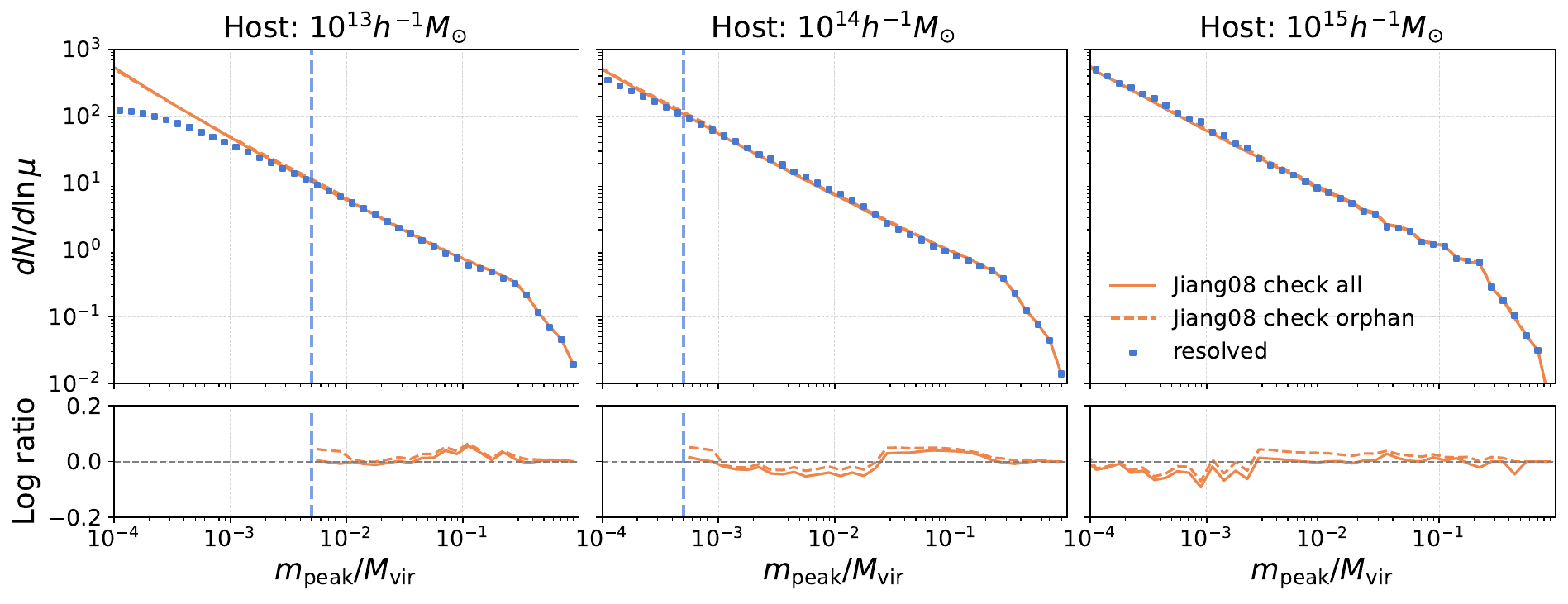}
    \caption{Comparison of SPMFs in FOF halos at $z=0$ from the combined \textsc{Jiutian-300} and \textsc{Jiutian-1G} results, compared with predictions based on the merger timescale model from Jiang et al. \cite{2008ApJ...675.1095J}. Solid lines show results using the model applied to all infall subhalos, while dashed lines show results when applying the model only to orphans, with all resolved subhalos retained. Results are shown for three host halo mass bins. The blue dashed vertical lines indicate the subhalo mass corresponding to 5000 particles in \textsc{Jiutian-300}.}
    \label{fig:compare_method}
\end{figure}

\section{Spatial distribution of surviving subhalos}\label{sec:space_dis}
With the SPMFs shown in figure~\ref{fig:evo_fof}, the total subhalo abundance can be accurately recovered by appropriately including a population of orphan subhalos. However, as mentioned above, there is no guarantee that the phase-space distribution is also unbiased. Therefore, I test and improve the commonly adopted method for handling orphans and examine whether the spatial distributions can be fully recovered.

\subsection{Orphan subhalo inclusion}\label{sec:orphan}
A common procedure for including orphan subhalos is to retain all resolved subhalos and consider only the merger timescales for the orphans. Orphans predicted to survive based on their merger timescales are included in the catalog, with their phase-space information traced using their most bound particles. However, merger timescale models are typically calibrated for the entire infall population rather than for orphans specifically. If orphans represent a biased subset in terms of merger timescales, applying these models exclusively to orphans may be inappropriate. I compare the SPMFs in FOF halos obtained by applying the merger timescale model to all infall subhalos versus only to orphans in figure~\ref{fig:compare_method}. The SPMFs from checking only orphans are typically 5–10\% higher than those from checking all infall subhalos, but this difference is smaller than the discrepancy between either model and the simulation results. Therefore, the primary limitation still lies in the merger timescale model itself, rather than in the choice between applying it to all infall subhalos or only to orphans. 

To make the inclusion of orphans more model-independent, I adopt a method that yields an exact match to the SPMFs shown in figure~\ref{fig:evo_fof}. In this approach, the total subhalo abundance depends on the merger timescale model only at the low-mass end ($m_{\rm peak} < 10^{10.7}\,h^{-1}M_{\odot}$), where even \textsc{Jiutian-300} becomes incomplete. First, I compute the ratio $T_{\rm ratio} = T_{\rm merger} / [T(z) - T_{\rm infall}]$ for each orphan subhalo, where $T(z)$ is the cosmic time at the redshift of interest and $T_{\rm infall}$ is the infall time of the orphan. Here I use a time ratio rather than a time difference because the scatter in $T_{\rm merger}$ around best-fit models is typically lognormal. Then, in each $M_{\rm vir}$ and $\mu$ bin, orphans are sorted by $T_{\rm ratio}$ in descending order, and those with the highest values are included until the total number matches the SPMFs shown in figure~\ref{fig:evo_fof}. With this method, only the relative values of $T_{\rm merger}$ matter at the high-mass end. In the case of a perfect merger timescale model, the inclusion would stop at orphans exactly with $T_{\rm ratio} = 1$. 

\begin{figure}[htbp]
    \centering
    \includegraphics[width=\textwidth]{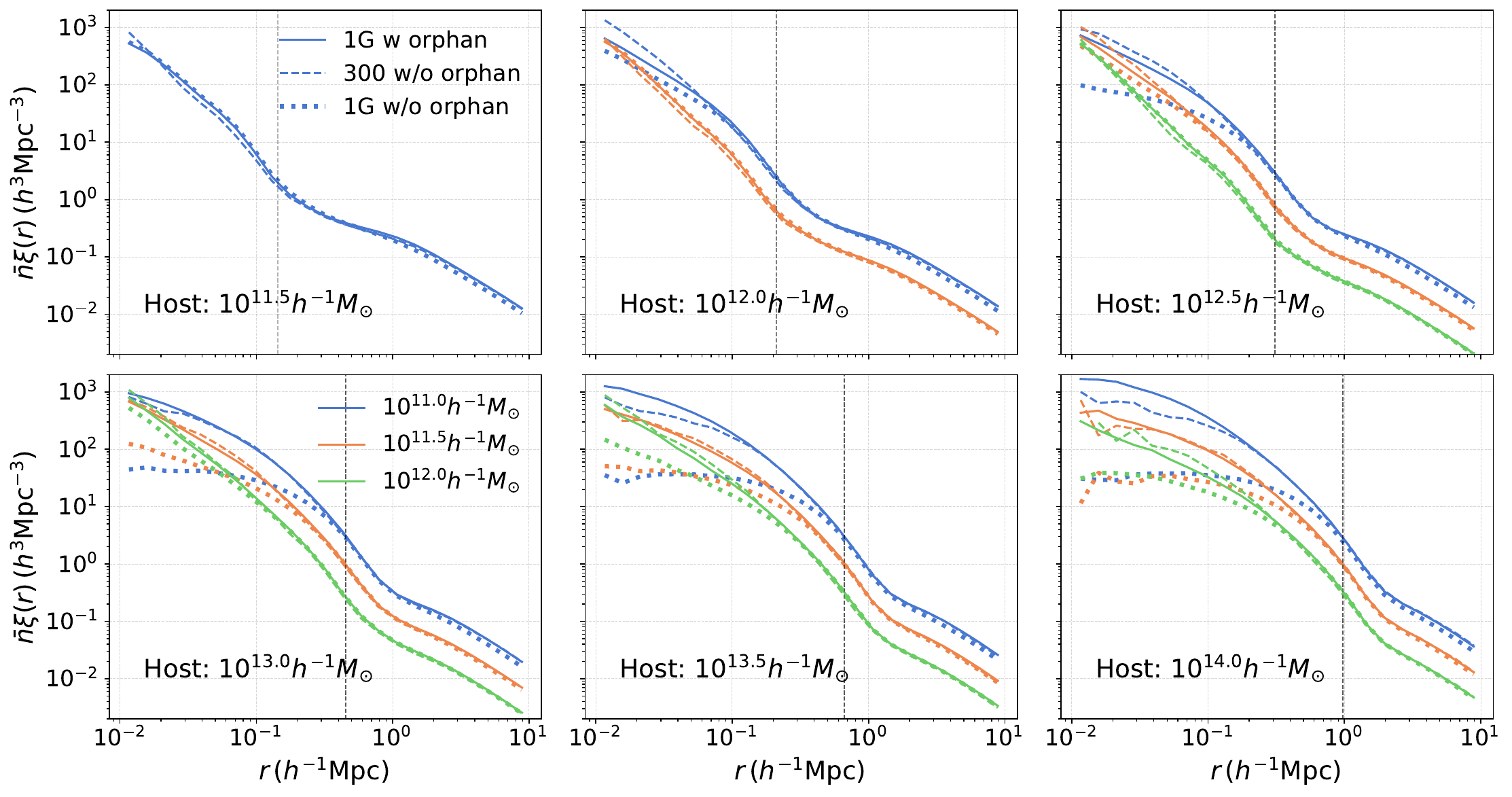}
    \caption{Spatial distribution of (sub)halos with specified masses around host halos in selected mass bins at $z=0.1$. Results are shown for neighboring objects with $10^{10.7} < m_{\rm peak} < 10^{12.3}\,h^{-1}M_{\odot}$, a range where subhalos in \textsc{Jiutian-300} are well resolved but subject to resolution limitations in \textsc{Jiutian-1G}. We compare results from \textsc{Jiutian-300}, \textsc{Jiutian-1G} without orphans, and \textsc{Jiutian-1G} with orphans included using the new method described in section~\ref{sec:orphan}. Vertical dashed lines indicate the location of $R_{\rm vir}$. }
    \label{fig:spatial_compare}
\end{figure}

\begin{figure}[htbp]
    \centering
    \includegraphics[width=\textwidth]{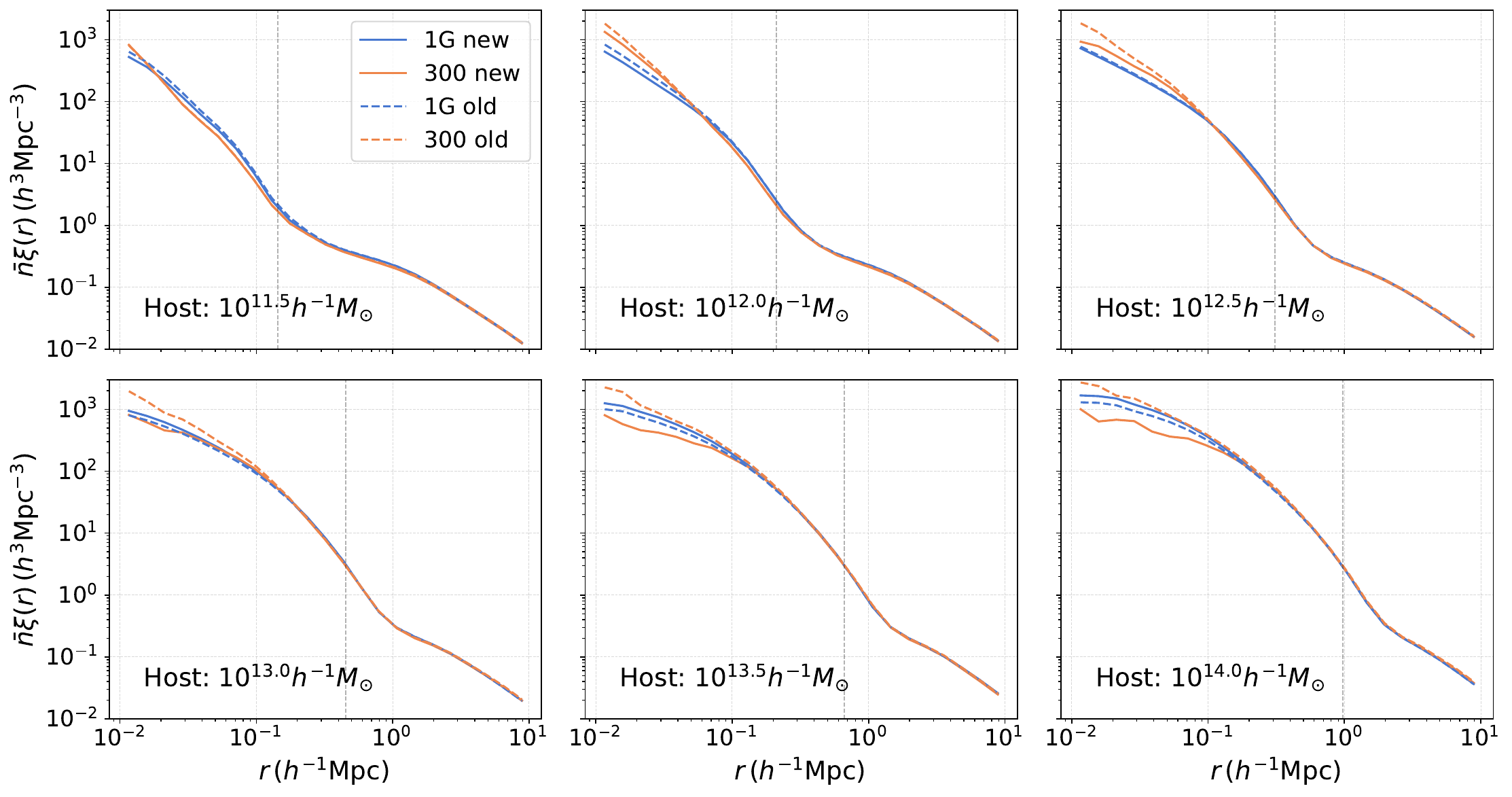}
    \caption{Similar to figure~\ref{fig:spatial_compare}, but comparing two methods for including orphans in the spatial distribution of neighboring objects with $m_{\rm peak} = 10^{11.0}\,h^{-1}M_{\odot}$. The new method refers to the SPMF-matching approach described in section~\ref{sec:orphan}, while the old method retains all resolved subhalos and includes orphans based on their individual merger timescales.}
    \label{fig:spatial_method}
\end{figure}

\subsection{Numerical convergence of spatial distribution}
I investigate whether the spatial distributions of subhalos can also be accurately recovered by including orphan subhalos, in addition to the total abundance. Figure~\ref{fig:spatial_compare} presents the excess density profiles $\bar{n}\xi(r)$ of (sub)halos around host halos in different mass bins, with subhalo masses defined by $m_{\rm peak}$ and host halo masses by $M_{\rm vir}$. Results are shown for \textsc{Jiutian-1G} and \textsc{Jiutian-300} without orphans, as well as for \textsc{Jiutian-1G} with orphans included using the SPMF-matching method introduced in section~\ref{sec:orphan}. Compared to \textsc{Jiutian-300}, figure~\ref{fig:spatial_method} shows an increasing deficit of subhalos in \textsc{Jiutian-1G} toward smaller $\mu$ and smaller radial distances $r$, with up to an order-of-magnitude loss of $10^{11.0}\,h^{-1}M_{\odot}$ subhalos in $10^{14.0}\,h^{-1}M_{\odot}$ halos. After including orphans in \textsc{Jiutian-1G}, the spatial distributions of subhalos are well recovered for $10^{11.5}\,h^{-1}M_{\odot}$ and $10^{12.0}\,h^{-1}M_{\odot}$ across all host halo masses and radial scales. For $10^{11.0}\,h^{-1}M_{\odot}$ subhalos at $r < 0.1\,h^{-1}\mathrm{Mpc}$, the distributions are largely recovered but still show noticeable deviations from \textsc{Jiutian-300}. These results suggest that, for low-mass subhalos in the innermost regions of halos, a more sophisticated method is required to accurately model the phase-space distribution of orphans. 

Without higher-resolution simulations, it is difficult to determine which result is more accurate in the unconverged region—\textsc{Jiutian-300} without orphans or \textsc{Jiutian-1G} with orphans. Although \textsc{Jiutian-300} appears to converge at these masses based on the SPMF, the inner regions of massive halos contribute only a very small fraction to the total abundance—possibly even smaller than the cosmic variance. Therefore, it remains possible that \textsc{Jiutian-300} is also incomplete in these regions. To further investigate this point, I also show the results for $10^{11.0}\,h^{-1}M_{\odot}$ in figure~\ref{fig:spatial_method}, using the old method that retains all resolved subhalos and includes orphans based on their individual merger timescales, and compare them to the results from the SPMF-matching method. While the old method may not reproduce the total abundance as accurately as the new one, it may better recover the spatial distribution, since orphans in \textsc{Jiutian-300} are also examined for this mass bin. However, in figure~\ref{fig:spatial_method}, I find that all four results from \textsc{Jiutian-1G} and \textsc{Jiutian-300}, using both the new and old orphan inclusion methods, disagree at $r < 0.1\,h^{-1}\mathrm{Mpc}$, making it difficult to determine which is more accurate. In fact, even in much higher resolution simulations ($m_{\rm p} \sim 10^{3}\,M_{\odot}$), convergence remains challenging in this region \cite{2025MNRAS.540.1107S}. Moreover, even if convergence in N-body simulations can be achieved, this inner region is significantly influenced by the gravitational and baryonic effects of central galaxies, which must be accounted for when modeling realistic data. Since this is precisely the region where many small-scale challenges in the Milky Way halo emerge \cite{1999ApJ...522...82K, 1999ApJ...524L..19M}, conclusions drawn from such models should be approached with caution. 

\section{Velocity distribution of surviving subhalos}\label{sec:velocity}
Velocity-based cosmological probes, such as satellite kinematics \cite{2012ApJ...758...50L} and redshift-space distortions (RSD) \cite{1987MNRAS.227....1K}, are powerful tools for studying structure formation. However, their inherently non-linear nature makes theoretical modeling particularly challenging. For instance, modeling the Fingers-of-God (FOG) effect in RSD typically involves assumptions about the galaxy velocity distribution, whether in perturbation theory or halo-based frameworks \cite{2001Natur.410..169P,2023ApJ...948...99Z,2024MNRAS.534.3595P}, introducing substantial uncertainties into the models. In contrast, when using subhalos, the velocity field can be more directly modeled through the galaxy–halo connection. For example, Gao et al. \cite{2023ApJ...954..207G,2024ApJ...961...74G} successfully predicted the redshift-space clustering of DESI emission line galaxies down to scales of $s > 0.3\,h^{-1}\mathrm{Mpc}$ using subhalo abundance matching. These models require an accurate characterization of the subhalo velocity field. Therefore, in this section, I investigate the numerical convergence of the velocity distribution in \textsc{Jiutian-1G} and \textsc{Jiutian-300}.

\begin{figure}[htbp]
    \centering
    \includegraphics[width=\textwidth]{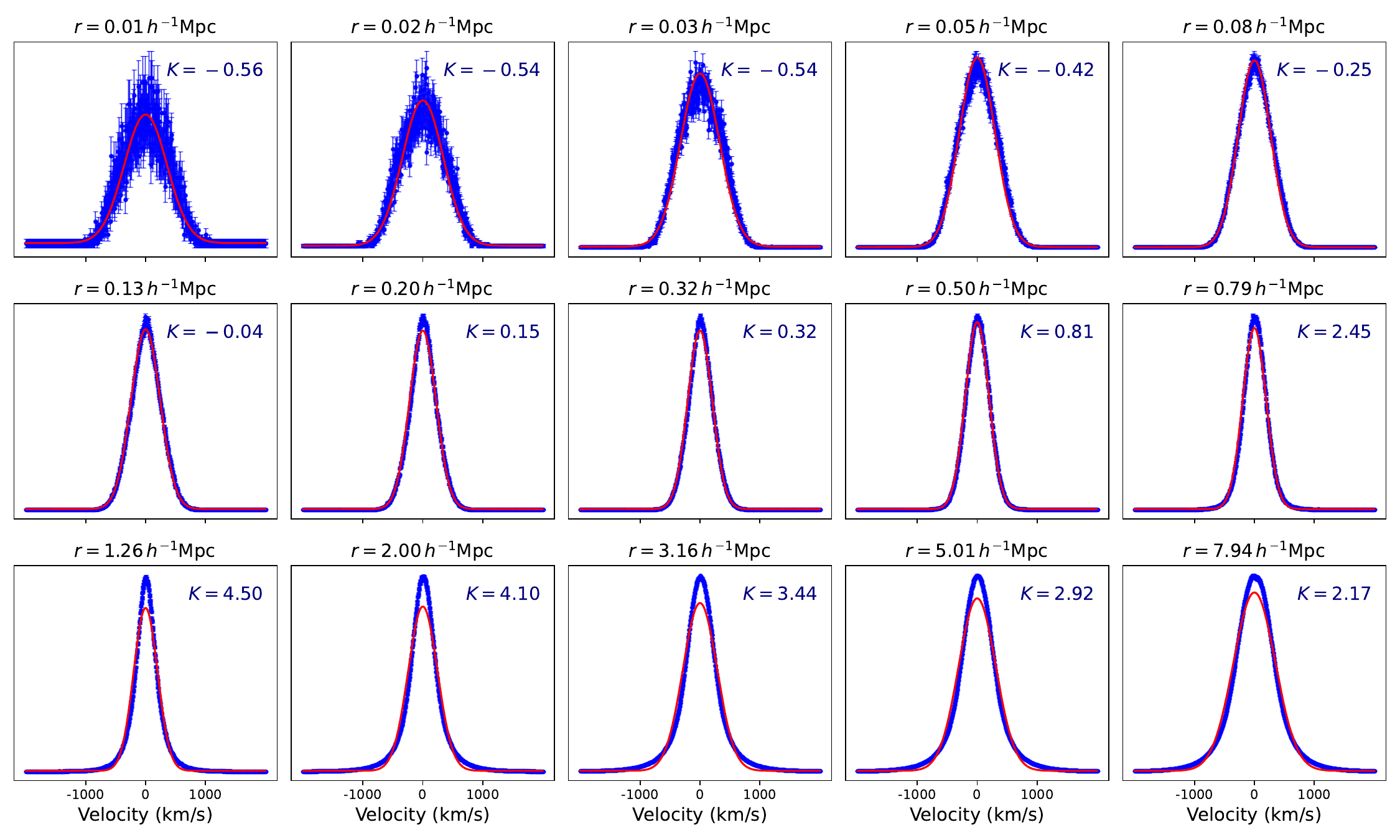}
    \caption{A fitting example of the residual $z$-direction velocity distributions in different radial bins for host halos of mass $10^{13.0}\,h^{-1}M_\odot$ and neighboring (sub)halos of $10^{11.0}\,h^{-1}M_\odot$. Blue dots with error bars represent measurements with Poisson uncertainties, while red lines show the best-fit Gaussian models. The excess kurtosis $K$ in each panel quantifies the deviation from Gaussianity.}
    \label{fig:vd_fit}
\end{figure}

\begin{figure}
    \centering
    \includegraphics[width=\textwidth]{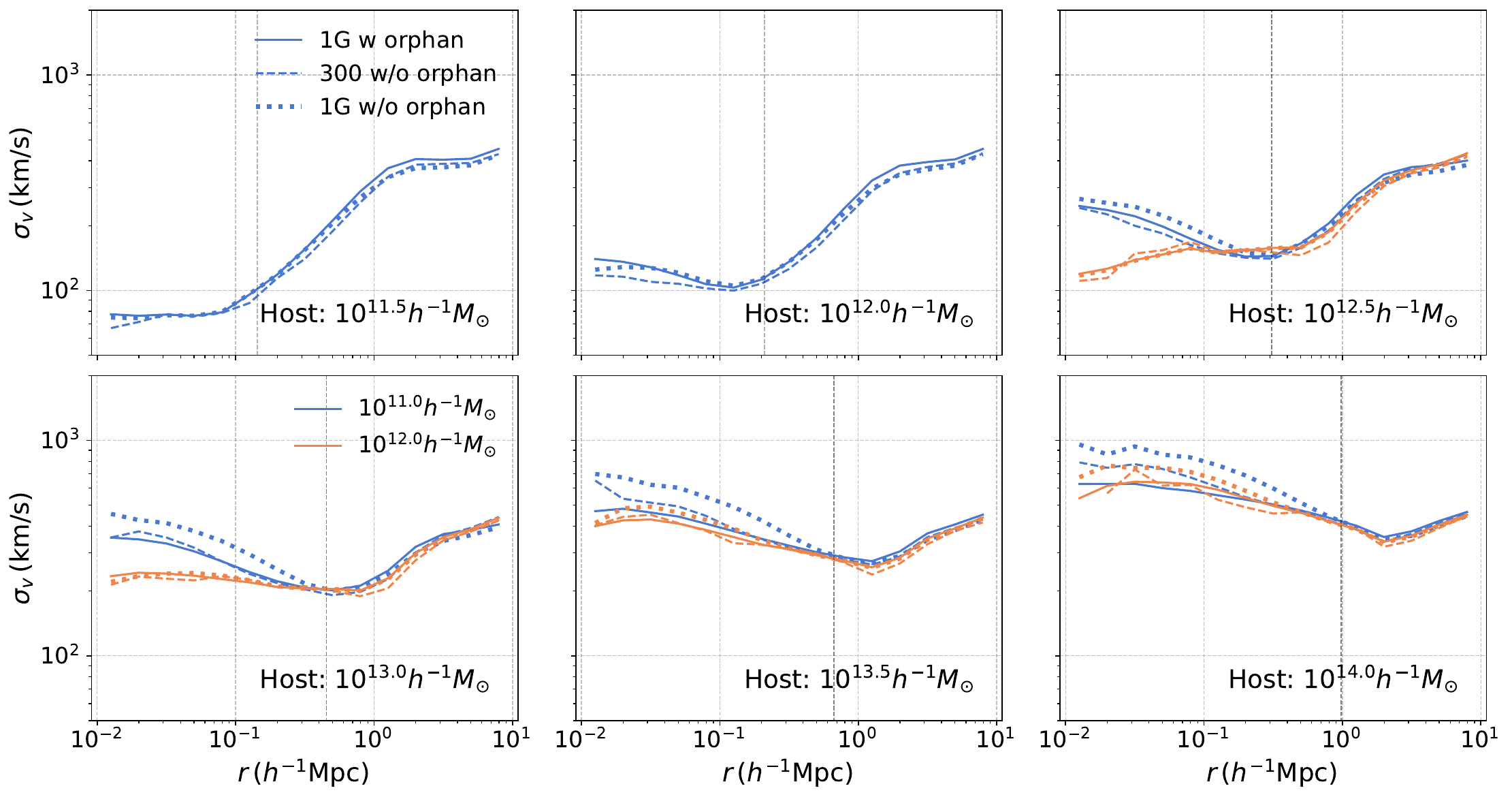}
    \caption{One-dimensional velocity dispersion $\sigma_{v}$ of (sub)halos with specified masses around host halos in selected mass bins at $z=0.1$. I compare results from \textsc{Jiutian-300}, \textsc{Jiutian-1G} without orphans, and \textsc{Jiutian-1G} with orphans included using the SPMF-matching method. Vertical dashed lines indicate the location of $R_{\rm vir}$.}
    \label{fig:vd_compare}
\end{figure}

\begin{figure}[htb]
    \centering
    \includegraphics[width=\textwidth]{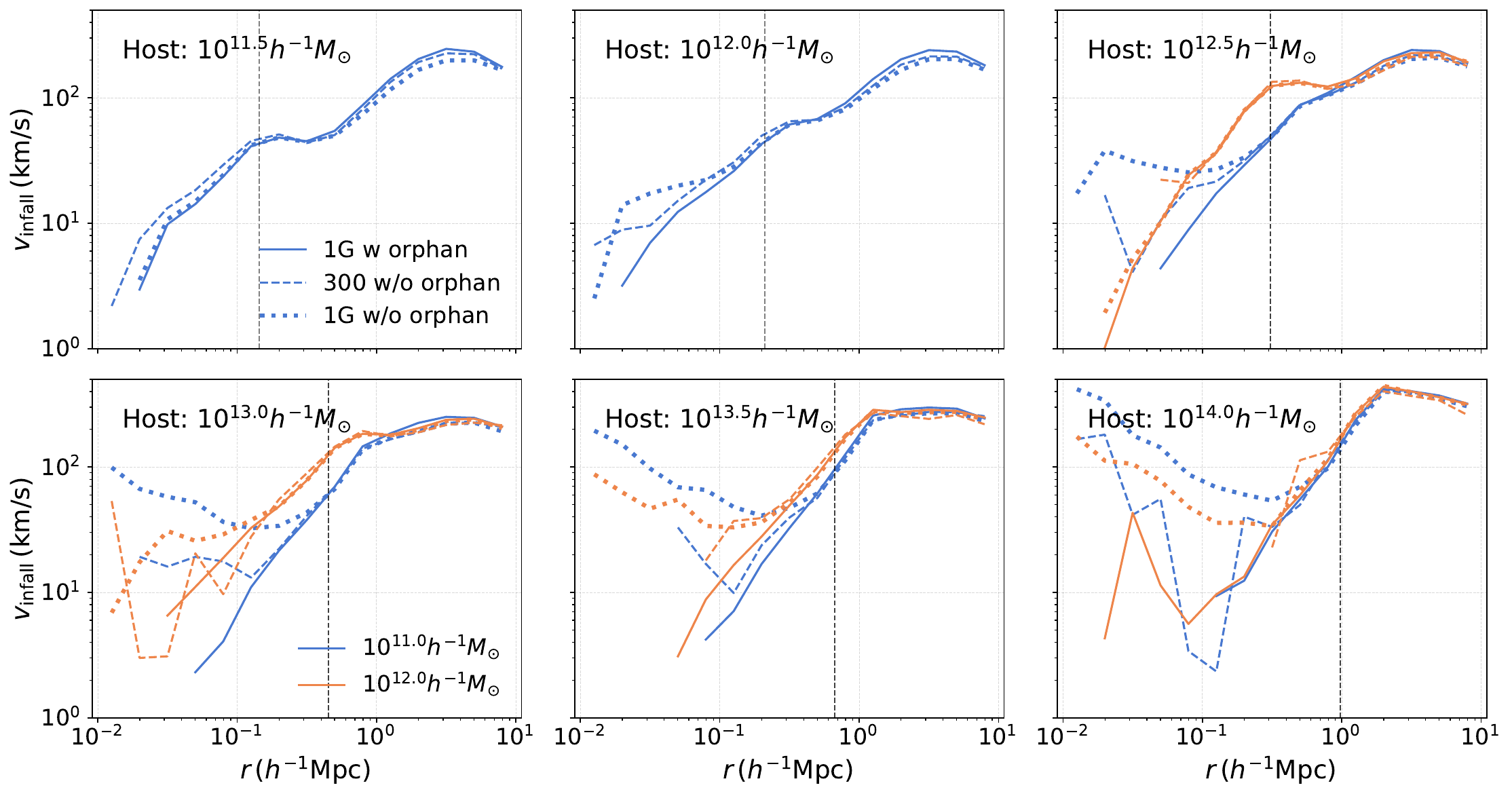}
    \caption{The average infall velocity $v_{\rm infall}$ of (sub)halos with specified masses around host halos in selected mass bins at $z=0.1$. We compare results from \textsc{Jiutian-300}, \textsc{Jiutian-1G} without orphans, and \textsc{Jiutian-1G} with orphans included using the SPMF-matching method. Vertical dashed lines indicate the location of $R_{\rm vir}$.}
    \label{fig:vinfall_compare}
\end{figure}

\begin{figure}
    \centering
    \includegraphics[width=\textwidth]{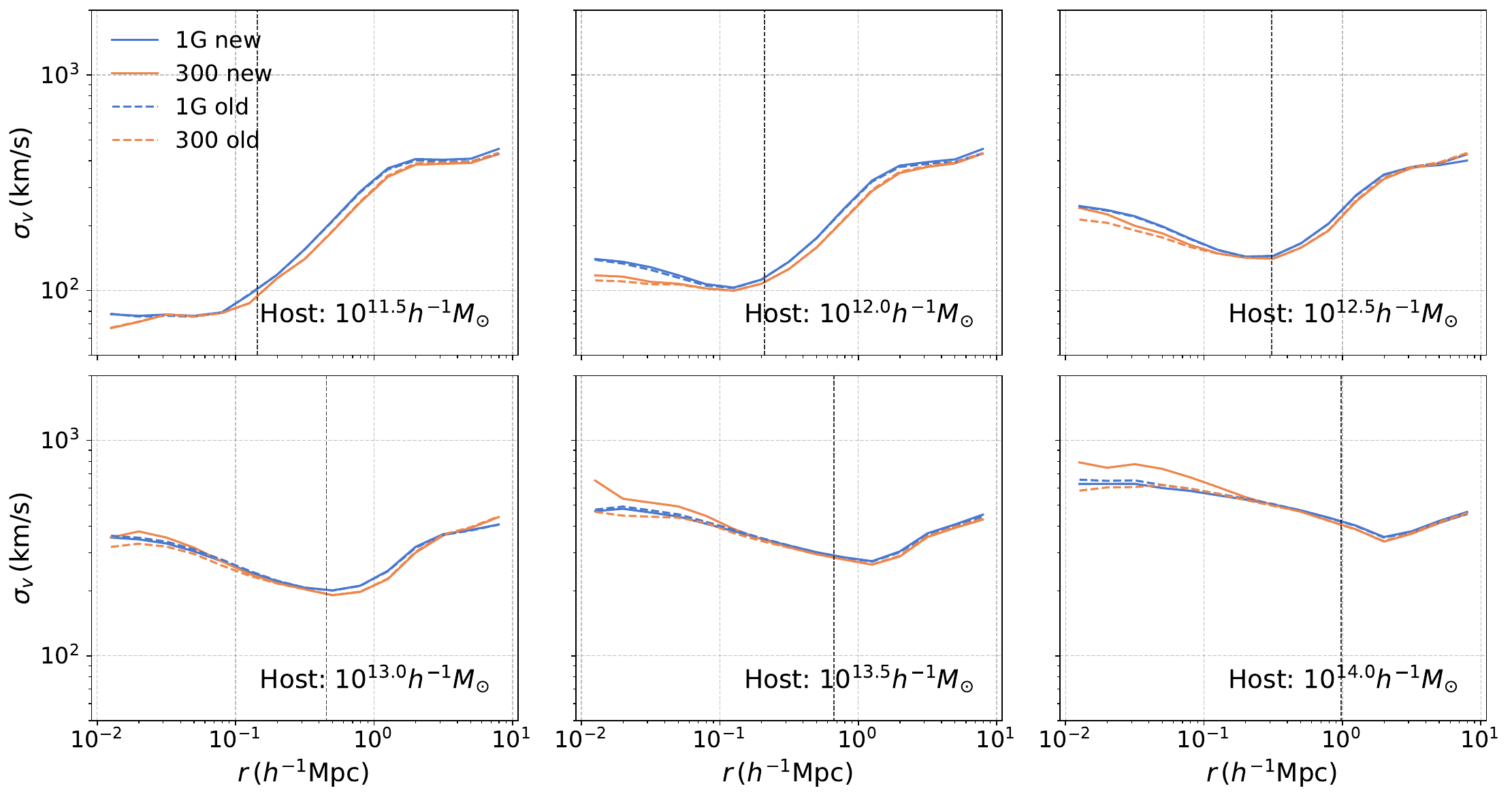}
    \caption{Similar to figure~\ref{fig:vd_compare}, but comparing two methods for including orphans in $\sigma_v$ of neighboring objects with $m_{\rm peak} = 10^{11.0}\,h^{-1}M_{\odot}$. The new method refers to the SPMF-matching approach described in section~\ref{sec:orphan}, while the old method retains all resolved subhalos and includes orphans based on their individual merger timescales.}
    \label{fig:vd_method}
\end{figure}

\subsection{Infall velocity and velocity dispersion}
Although the total line-of-sight (LOS) peculiar velocity is what directly affects RSD and other velocity-based cosmological probes, it is more instructive to understand the subhalo velocity field by decomposing the total velocity into radial infall velocity ($v_{\rm infall}$) and velocity dispersion ($\sigma_v$) components. In this paper, $\sigma_v$ refers to the one-dimensional velocity dispersion, while the total velocity dispersion is given by $\sqrt{3}\sigma_v$ for an isotropic system. In an isolated, virialized system, the velocity dispersion fully characterizes the total velocity field. However, in a hierarchically forming Universe, where overdense regions continuously accrete nearby smaller structures, a non-zero radial infall velocity is required to account for the influence of large-scale bulk motions.

Following the scheme in section~\ref{sec:space_dis}, I measure the average radial velocity and velocity dispersion of small (sub)halos around host halos in different mass bins over the range $0.01 < r < 10\,h^{-1}{\rm Mpc}$. After subtracting the infall velocity, the velocity dispersion $\sigma_v$ is determined either by directly computing the variance or by fitting a Gaussian profile. These two methods should yield consistent results if the distribution of residual velocities is Gaussian. In figure~\ref{fig:vd_fit}, I show an example of Gaussian fitting for host halos of mass $10^{13.0}\,h^{-1}M_\odot$ and neighboring (sub)halos of $10^{11.0}\,h^{-1}M_\odot$. Within $R_{\rm vir}$, the residual velocities along the $z$-direction are well described by a Gaussian distribution. However, at larger scales, the distributions exhibit leptokurtic features, characterized by heavier tails and a sharper peak. To quantify this non-Gaussianity, I calculate the excess kurtosis $K$ of each distribution, defined as
\begin{equation}
K = \frac{\mathbb{E}[(X - \mu)^4]}{\sigma^4} - 3,
\end{equation}
where $\mu$ and $\sigma$ are the mean and standard deviation of the distribution, respectively. A Gaussian distribution has $K = 0$; $K > 0$ indicates a leptokurtic deviation (heavier tails and sharper peak), while $K < 0$ corresponds to a platykurtic deviation (lighter tails and flatter peak). The non-Gaussian nature of the velocity distribution at large scales may arise from the fact that subhalos at these scales reside in host halos with a wide range of $M_{\rm vir}$. Since $\sigma_v(r)$ is strongly dependent on $M_{\rm vir}$, the resulting distribution is effectively a superposition of Gaussian distributions with varying $\sigma_v(r)$, which does not yield a Gaussian overall. In contrast, within $R_{\rm vir}$, where halos are binned by $M_{\rm vir}$, the stacked distributions have similar $\sigma_v(r)$ and thus retain a highly Gaussian shape. Due to the non-Gaussian feature at large scale, $\sigma_v$ below is calculated by directly computing the variance. Although $\sigma_v$ alone does not contain all the information of the velocity distribution at large scale, it is still a good quantity for the test of numerical convergence. 

In figures~\ref{fig:vd_compare} and \ref{fig:vinfall_compare}, I compare $\sigma_v$ and $v_{\rm infall}$ across different host halo masses and neighboring smaller (sub)halo mass bins for \textsc{Jiutian-1G} and \textsc{Jiutian-300} without orphans, as well as for \textsc{Jiutian-1G} with orphans included using the SPMF-matching method. A comparison between \textsc{Jiutian-1G} and \textsc{Jiutian-300} without orphans reveals that \textsc{Jiutian-1G} exhibits higher $\sigma_v$ and $v_{\rm infall}$. This suggests that subhalos with lower velocities are more susceptible to disruption. Such subhalos typically reside on inner orbits, spending more time near the center of their host halos. As a result, they experience stronger tidal heating over longer durations and are more sensitive to numerical effects. After incorporating orphans into \textsc{Jiutian-1G}, $\sigma_v$ is well-recovered down to $0.1\,h^{-1}{\rm Mpc}$, though it remains slightly higher (by 5\%–10\% for $10^{11.0}\,h^{-1}M_{\odot}$ subhalos) compared to \textsc{Jiutian-300}. This discrepancy arises because resolved subhalos experience dynamical friction, while most bound particles do not, leading to systematically higher particle velocities relative to the subhalos. It is worth noting that for the two largest host mass bins, \textsc{Jiutian-300} exhibits higher $\sigma_v$ than \textsc{Jiutian-1G} with correction for $10^{11.0}\,h^{-1}M_{\odot}$ subhalos at $r<0.1\,h^{-1}{\rm Mpc}$. This suggests that subhalos in \textsc{Jiutian-300} may already be affected by numerical effects, preferentially losing low-velocity subhalos in this region—even though the total abundance appears converged for this mass. To investigate further, I performed an analysis similar to the spatial distribution test: Figure~\ref{fig:vd_method} shows $\sigma_v$ results from the old subhalo inclusion method, which retains all resolved subhalos and incorporates orphans based on their individual merger timescales. After including orphans in \textsc{Jiutian-300}, $\sigma_v$ now converges with \textsc{Jiutian-1G} for these mass bins, confirming that numerical effects play a significant role in this regime. However, this convergence does not necessarily imply that the old method yields the correct $\sigma_v$. The agreement could instead reflect convergence toward the particle $\sigma_v$ rather than the true resolved subhalo $\sigma_v$, given the substantial fraction of orphans included in this region. Similar to the spatial distribution, accurately recovering the velocity distribution of subhalos within $r<0.1\,h^{-1}{\rm Mpc}$ remains challenging when using the simple orphan inclusion method, necessitating a more sophisticated treatment for this region.

From figures~\ref{fig:vd_compare}, I observe that $\sigma_v$ exhibits a V-shaped radial profile, reaching its minimum near $R_{\rm vir}$ - likely corresponding to the splashback radius \cite{2014ApJ...789....1D}. This characteristic profile may arise from halo exclusion effects: at small radii, $\sigma_v$ primarily reflects the host halo's internal velocity dispersion, while at large radii it represents the average $\sigma_v$ of neighboring halos. The minimum near $R_{\rm vir}$ corresponds to the infall region, where subhalos experience coherent accretion flows with reduced velocity dispersion. Furthermore, figure~\ref{fig:vinfall_compare} reveals that $v_{\rm infall}$ increases toward smaller scales from large radii, peaking at a few times $R_{\rm vir}$ before decreasing due to virialization. This behavior is consistent with previous findings by \cite{1999MNRAS.305..547C}. The large-scale trend can be understood through linear theory \cite{1980lssu.book.....P,1999MNRAS.305..547C}:
\begin{equation}
v_{\rm infall}^{\rm lin}(r)=-\frac{1}{3}H_0\Omega_m^{0.6}r\delta(<r)\,,
\end{equation}
where $\delta(<r)$ represents the mean overdensity within radius $r$.

It is also interesting to compare the results for subhalos of different masses in figures~\ref{fig:vd_compare} and \ref{fig:vinfall_compare}. For $\sigma_v$, more massive subhalos exhibit smaller values, a consequence of stronger dynamical friction acting on them. The difference between the two subhalo mass bins becomes less pronounced in more massive host halos, as the dynamical friction for both samples is reduced due to their smaller mass ratios. This also explains the difference in spatial distributions seen in figure~\ref{fig:spatial_compare}, where more massive subhalos exhibit steeper inner profiles, as a larger fraction of them sink toward the center under the influence of stronger dynamical friction. In contrast, $v_{\rm infall}$ exhibits a different trend, with more massive subhalos showing larger $v_{\rm infall}$ values within the host halo. This is because larger subhalos tend to have more eccentric, radially biased orbits \cite{2015MNRAS.448.1674J}, as they are less easily virialized due to being less perturbed by smaller subhalos.

\begin{figure}
    \centering
    \includegraphics[width=\textwidth]{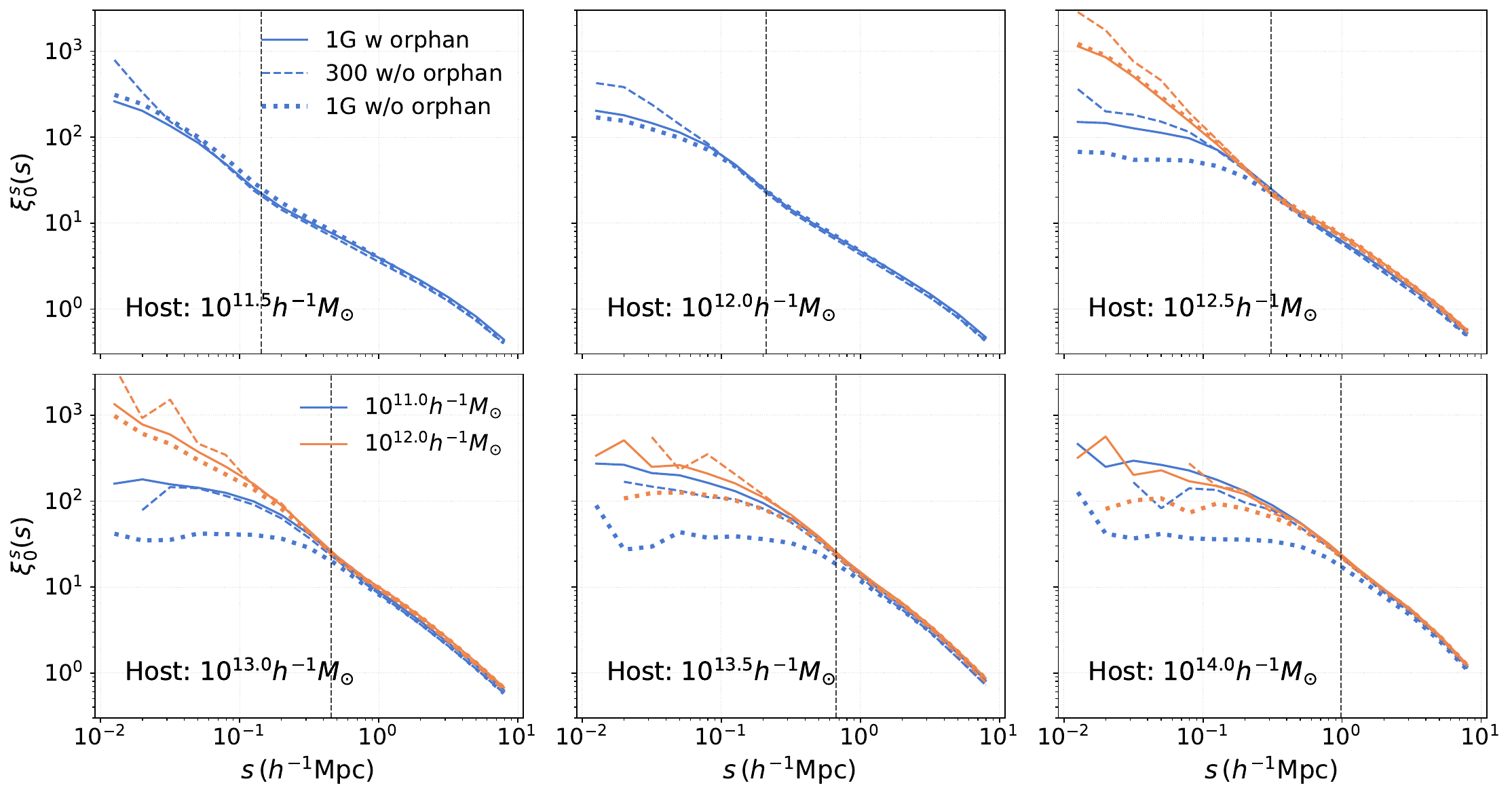}
    \caption{Redshift-space two-point correlation function monopoles $\xi^s_0(s)$ for (sub)halos of specified mass ranges, measured around host halos in selected mass bins at redshift $z=0.1$. I compare results from \textsc{Jiutian-300}, \textsc{Jiutian-1G} without orphans, and \textsc{Jiutian-1G} with orphans included using the SPMF-matching method. Vertical dashed lines indicate the location of $R_{\rm vir}$.}
    \label{fig:xi0}
\end{figure}

\begin{figure}
    \centering
    \includegraphics[width=\textwidth]{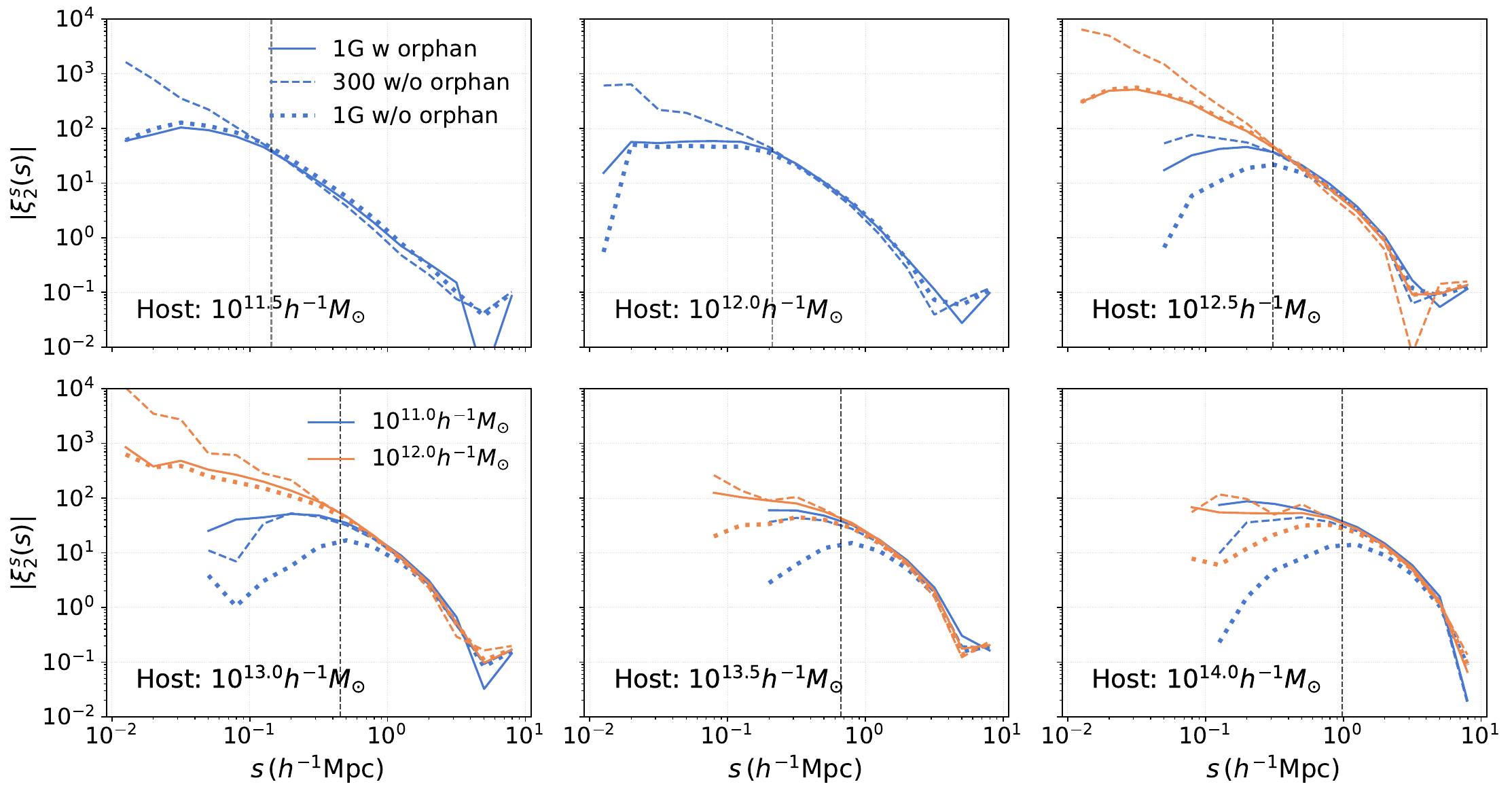}
    \caption{Similar to figure~\ref{fig:xi0}, but showing the absolute value of the redshift-space quadrupoles, $|\xi^s_2(s)|$.}
    \label{fig:xi2}
\end{figure}

\begin{figure}
    \centering
    \includegraphics[width=\textwidth]{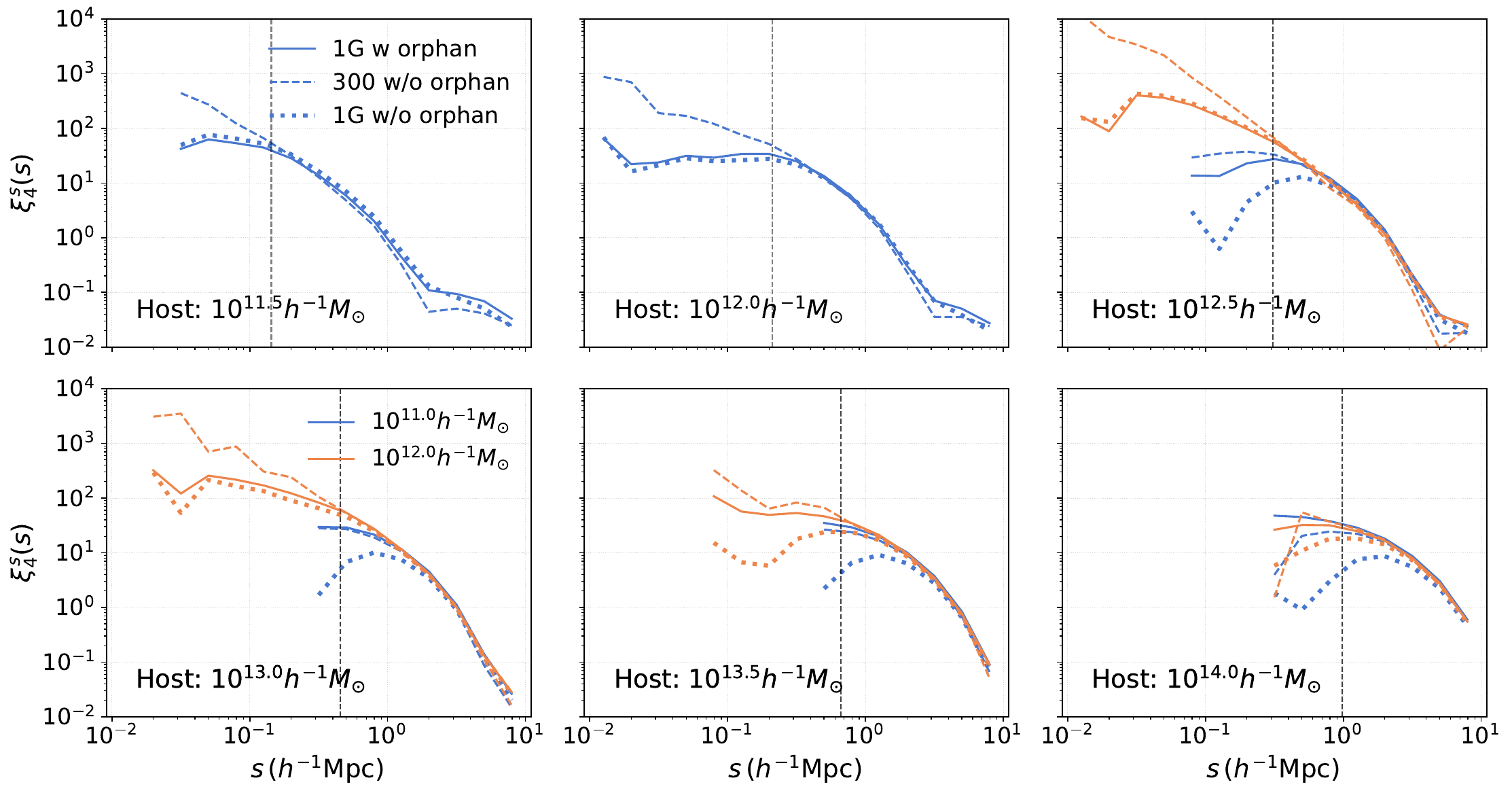}
    \caption{Similar to figure~\ref{fig:xi0}, but showing the redshift-space hexadecapoles, $\xi^s_4(s)$.}
    \label{fig:xi4}
\end{figure}

\subsection{Redshift-space clustering}
Although degenerate with the Hubble flow, redshift remains the only observable to probe the cosmic peculiar velocity field, making galaxy clustering in redshift space an important and powerful cosmological probe. Therefore, I further investigate the convergence of the mock redshift-space clustering of halos and subhalos, which depends on the combined effects of their real-space spatial and velocity distributions.

I translate the positions of each halo and subhalo from real space, $\mathbf{r}$, to redshift space, $\mathbf{s}$, through
\begin{equation}
    \mathbf{s} = \mathbf{r} + \frac{v_{\parallel}}{aH} \hat{\mathbf{n}}\,,
\end{equation}
where $v_{\parallel}$ is the LOS peculiar velocity, $a$ is the scale factor, $H$ is the Hubble parameter, and $\hat{\mathbf{n}}$ is the LOS direction chosen as the $z$-direction in the simulations. The statistic containing the full two-point information is the two-dimensional correlation function, $\xi^s(r_{\rm p}, \Pi)$, where $r_{\rm p}$ is the separation perpendicular to the LOS and $\Pi$ is the separation along the LOS. However, in many studies, due to limited data for two-dimensional binning, multiple moments of $\xi^s(r_{\rm p}, \Pi)$ are instead measured as  
\begin{equation}
    \xi^s_l(s) = \frac{2l+1}{2} \int_{-1}^{1} \xi^s(s,\mu) P_l(\mu) \, \mathrm{d}\mu \,,
\end{equation}
where $\mu = \Pi / s$ is the cosine of the angle between the separation vector and the LOS, and $P_l(\mu)$ is the Legendre polynomial of order $l$.

Since including orphans can recover both the spatial and velocity distributions at $r > 0.1\,h^{-1}{\rm Mpc}$, it is intuitive that $\xi^s(r_{\rm p}, \Pi)$ should also converge for $r_{\rm p} > 0.1\,h^{-1}{\rm Mpc}$ by definition. However, this is not the case for $s > 0.1\,h^{-1}{\rm Mpc}$. Owing to the elongated FOG effect, pairs with $r < 0.1\,h^{-1}{\rm Mpc}$ can extend to separations of tens of $h^{-1}{\rm Mpc}$ in $\mathbf{s}$-space, propagating numerical effects to very large scales. Although this issue can be mitigated by using a modified $\xi_l^s(s)$ that masks small $r_{\rm p}$ values or by adopting alternative statistics, $\xi_l^s(s)$ remains one of the most commonly used measures. It is therefore worth examining the scales at which caution is required, even after including orphan subhalos.

Figures~\ref{fig:xi0}, \ref{fig:xi2}, and \ref{fig:xi4} present comparisons of the redshift-space clustering multipoles - the monopole $\xi^s_0(s)$, quadrupole $\xi^s_2(s)$, and hexadecapole $\xi^s_4(s)$ - between three configurations: \textsc{Jiutian-300}, \textsc{Jiutian-1G} without orphans, and \textsc{Jiutian-1G} with orphans incorporated via the SPMF-matching method. While \textsc{Jiutian-1G} with orphans reproduces \textsc{Jiutian-300}'s real-space $\bar{n}\xi(r)$ and $\sigma_v$ to within $\sim5\%$ accuracy down to $0.1$–$0.2\,h^{-1}\mathrm{Mpc}$ for $10^{11.0}\,h^{-1}M_{\odot}$ subhalos, the redshift-space multipoles match only at considerably larger scales. The monopole $\xi^s_0(s)$ reaches approximate agreement at scales about twice those required for real-space statistics, whereas the quadrupole $\xi^s_2(s)$ and hexadecapole $\xi^s_4(s)$ converge more gradually, achieving better consistency only beyond $R_{\rm vir}$. Even at these scales, $\xi^s_0(s)$ still shows a difference of about $10\%$, while $\xi^s_2(s)$ and $\xi^s_4(s)$ remain more than $20\%$ apart. These trends are expected as a consequence of the elongated FOG effect discussed earlier. Consequently, it may be preferable to adopt modified or alternative redshift-space statistics for subhalo-based modeling—such as introducing a maximum $\mu$ cutoff in the multipole integration to exclude pairs with small transverse separations $r_{\rm p}$. In this case, the modified multipoles are defined as  
\begin{equation}
    \xi_{l,{\rm mod}}^s = \frac{2l+1}{2} \int_{-{\mu^{\rm max}}}^{{\mu^{\rm max}}} \xi^s(s,\mu) \, P_l(\mu) \, \mathrm{d}\mu \,,
\end{equation}
where $\mu^{\rm max} = \sqrt{s^2 - (r_{\rm p}^{\rm min})^2} / s$ and $r_{\rm p}^{\rm min}$ is the minimum transverse separation included. I do not explore this further, as the choice is somewhat arbitrary and the outcomes are largely intuitive given the $\bar{n}\xi(r)$ and $\sigma_v$ results.

\section{Conclusion}\label{sec:con}
In this paper, I study the numerical convergence and correction methods of the abundance, spatial distribution, and velocity distribution of subhalos using two simulations with different mass resolutions, \textsc{Jiutian-300} and \textsc{Jiutian-1G}. Meanwhile, my results also contribute to the study and understanding of these fundamental properties. The main results are summarized as follows:
\begin{itemize}
    \item The SPMF converges only at $m_{\rm peak}$ corresponding to 5000 particles, and it can be accurately recovered by including orphan subhalos that are still surviving according to the merger timescale model of Jiang08 \cite{2008ApJ...675.1095J}, which outperforms other models.
    \item I construct the SPMF by combining the two simulations and the Jiang08 model over a wide range of host halo masses ($10^{10.0} < M_{\rm vir} < 10^{15.5},h^{-1}M_{\odot}$), mass ratios ($10^{-6} < \mu < 1$), and redshifts ($0 < z < 5$), and provide a fitting formula that accurately (within $\sim$10\%) captures all these dependencies.

    \item The real-space spatial ($\bar{n}\xi(r)$) and velocity ($\sigma_v$) distributions can be recovered with $5\%-10\%$ accuracy down to scales of $0.1$–$0.2\,h^{-1}{\rm Mpc}$ by including orphan subhalos, with more massive subhalos and lower-mass host halos showing better agreement at larger scales. The remaining differences are likely due to cosmic variance and finite-box effects in the smaller \textsc{Jiutian-300} simulation. Convergence at scales below $0.1\,h^{-1}{\rm Mpc}$ is difficult to achieve and requires more sophisticated modeling of orphan subhalos.

    \item Within the same host halos, more massive subhalos exhibit smaller velocity dispersion ($\sigma_v$) and steeper density profiles due to stronger dynamical friction. They also have larger infall velocities ($v_{\rm infall}$) resulting from more eccentric, radially biased orbits.

    \item As a consequence of the elongated FOG effect, redshift-space multipoles $\xi^s_l(s)$ are harder to recover even with orphan subhalos because unreliable small-scale pairs at $r_{\rm p} < 0.1\,h^{-1}{\rm Mpc}$ in real space can affect scales of tens of $h^{-1}$ Mpc in redshift space. Therefore, using modified or alternative redshift-space statistics that minimize the influence of small $r_{\rm p}$ is recommended for subhalo-based modeling.
\end{itemize}

This work only studies the simplest treatment of orphan subhalos by checking their survival and using the most bound particle’s position and velocity as their phase-space information. Although more sophisticated approaches are currently being discussed within the community, no widely accepted next-level treatment has yet emerged. In this sense, despite its simplicity, these results remain important as a starting point for understanding numerical effects on subhalo properties.
 

\acknowledgments
K.X. is supported by the funding from the Center for Particle Cosmology at U Penn. I thank Jiaxin Han, Carlos Frenk, Shaun Cole, and Yipeng Jing for valuable discussions, and Xiaolin Luo for assistance with downloading and organizing halo and subhalo catalogs. This work made use of the Gravity Supercomputer at the Department of Astronomy, Shanghai Jiao Tong University.



\bibliographystyle{JHEP}
\bibliography{biblio.bib}






\end{document}